\documentclass[aps,preprint,showpacs,showkeys,nofootinbib,floatfix,tightenlines]{revtex4}

\usepackage{ulem}
\usepackage{graphicx}  %
\usepackage{amsmath}
\usepackage{bm}  %

\newcommand{\bea}{\begin{eqnarray}}
\newcommand{\eea}{\end{eqnarray}}
\newcommand{\beq}{\begin{equation}}
\newcommand{\eeq}{\end{equation}}
\newcommand{\bqa}{\begin{eqnarray}}
\newcommand{\eqa}{\end{eqnarray}}

\def\mqo2{{\!\!\!}}

\begin{document}

\title{
Avalanche Mechanism \\
for the Enhanced Loss 
of Ultracold Atoms}

\author{Christian Langmack}
\affiliation{Department of Physics,
         The Ohio State University, Columbus, OH\ 43210, USA}

\author{D.~Hudson Smith}
\affiliation{Department of Physics,
         The Ohio State University, Columbus, OH\ 43210, USA}

\author{Eric Braaten}
\affiliation{Department of Physics,
         The Ohio State University, Columbus, OH\ 43210, USA}

\date{\today}

\begin{abstract}
In several experiments with ultracold trapped atoms,
a narrow loss feature has been observed 
near an {\it atom-dimer resonance}, at which there is
an Efimov trimer at the atom-dimer threshold. 
The conventional interpretation of these loss features is
that they are produced by the {\it avalanche mechanism},
in which the energetic atom and dimer from 3-body recombination
undergo secondary elastic collisions that produce additional atoms 
with sufficient energy to escape from the trapping potential.
We use Monte Carlo methods to calculate the average number of atoms lost 
and the average heat generated by recombination events in a 
Bose-Einstein condensate and in a thermal gas.
We improve on previous models by taking into account the energy-dependence 
of the cross sections, the spacial structure of the atom cloud,
and the elastic scattering of the atoms.  
We show that the avalanche mechanism cannot produce a narrow loss feature 
near the atom-dimer resonance.
The number of atoms lost from a recombination event can be more than 
twice as large as the 3 that would be obtained in the absence of
secondary collisions.  However the resulting loss feature 
is broad and its peak is at a scattering length that is 
larger than the atom-dimer resonance and depends on the trap depth.
\end{abstract}

\smallskip
\maketitle

\section{Introduction}

Particles with short-range interactions and an S-wave scattering length $a$
that is large compared to the range have universal low-energy properties 
that depend on $a$ but not on other details of the interactions 
or on the structure of the particles \cite{Braaten:2004rn}.
This universality provides deep connections between various fields
of physics, including
atomic and molecular, condensed matter, nuclear, and particle physics.
It has stimulated dramatic advances in theoretical and experimental 
few-body physics -- particularly in the
study of the universal few-body reaction rates of ultracold atoms.

Since particles with large scattering length are essentially indivisible 
at low energies, we refer to them as {\it atoms}.
In the 2-atom sector, the universal properties are simple. 
If $a>0$, they include the existence of a loosely-bound diatomic molecule 
that we refer to as the {\it shallow dimer}.
In the 3-atom sector, the universal properties are more intricate. 
In many cases, including identical bosons,
they include the existence of a sequence of universal triatomic molecules
called {\it Efimov trimers} that were discovered by Efimov in 1970 \cite{Efimov70}.
In the zero-range limit, the spectrum of Efimov trimers 
is invariant under discrete scale transformations \cite{Efimov73}.
For identical bosons, the discrete scaling factor is approximately 22.7.
Reaction rates among three low-energy atoms also respect 
discrete scale invariance \cite{Efimov79}.
We refer to universal few-body phenomena with discrete scaling behavior 
as {\it Efimov physics}.

Ultracold trapped atoms provide an ideal laboratory for studying 
Efimov physics,  because the scattering length 
can be controlled experimentally using Feshbach resonances.
The simplest probes of Efimov physics 
are loss features:  local maxima and minima in the atom loss rate 
as functions of the scattering length $a$.
The most dramatic signature of an Efimov trimer is the resonant
enhancement of the 3-body recombination rate near a negative value of $a$
for which there is an Efimov trimer at the 3-atom threshold \cite{EGB-99}.
The first observation of such a loss feature in 
an ultracold gas of $^{133}$Cs atoms \cite{Grimm:0512} 
revealed a line shape consistent with 
universal predictions \cite{Braaten:2003yc}.

In a mixture of atoms and shallow dimers, a narrow loss feature 
can also be caused by an Efimov trimer near the atom-dimer threshold.
We refer to a scattering length $a_*$ for which an Efimov trimer
is exactly at the threshold as an {\it atom-dimer resonance}.
For $a$ near $a_*$, there is resonant enhancement near threshold of the
elastic scattering of an atom and the shallow dimer.  
There is also resonant enhancement of their inelastic scattering 
into an atom and a more tightly-bound diatomic molecule,
which we refer to as a {\it deep dimer}.  
The release of the large binding energy of the deep dimer gives
the outgoing atom and dimer large enough 
kinetic energies to escape from the trapping potential. 
The resulting peaks in the atom and dimer loss rates near $a_*$ 
were first observed 
in a mixture of $^{133}$Cs atoms and dimers \cite{Grimm:0807}.

There have also been observations of narrow loss features 
near an atom-dimer resonance in systems consisting of atoms only.
Zaccanti et al.\ observed a narrow loss peak near the predicted position 
of $a_*$ in a Bose-Einstein condensate 
of $^{39}$K atoms \cite{Zaccanti:0904}.
They also observed a loss peak in a thermal gas 
near the next atom-dimer resonance, at a scattering length larger 
by a factor of about 22.7.
Pollack et al.\ observed a loss peak near the predicted position 
of an atom-dimer resonance in a Bose-Einstein condensate 
of $^{7}$Li atoms \cite{Hulet:0911}.
Machtey et al.\ observed such a loss peak 
in a thermal gas of $^{7}$Li atoms \cite{Khaykovich:1201}.
These loss features near the atom-dimer resonance are puzzling, 
because the equilibrium population of shallow dimers is expected to be 
negligible in these systems.   
Thus, any losses due to inelastic scattering 
between an atom and a shallow dimer should be negligible.

Zaccanti et al.\ proposed an {\it avalanche mechanism} 
for the enhancement of the atom loss rate near an atom-dimer resonance
in systems consisting of atoms only \cite{Zaccanti:0904}.
Near $a_{*}$, atom-dimer cross sections are resonantly enhanced
near threshold.
Each 3-body recombination event produces an atom and a shallow dimer 
with kinetic energies much larger than that required to escape from the trap.
If the atom and dimer both escape, 3 atoms are lost. 
If the dimer instead scatters inelasticly, 
the scattered atom is also lost, so the number of atoms lost is 4.  
However the dimer can undergo one or more elastic collisions 
before ultimately escaping or suffering an inelastic collision,
and the scattered atoms may gain enough energy to escape from the trap.
The scattered atoms may also undergo elastic collisions, 
producing still more lost atoms.
Thus the dimer could initiate an avalanche of lost atoms.
The atom from the recombination event could also initiate 
an avalanche of lost atoms.
Thus the number of atoms lost could be significantly larger than 3.
Near $a_*$, the resonant enhancement 
of the atom-dimer elastic cross section increases the 
probability for the dimer to undergo an elastic collision
and initiate an avalanche.
This suggests that there should be an increase in the number 
of atoms lost per recombination event
near $a_*$.  If the increase is sufficiently narrow, 
it could be observed as a local maximum in the atom loss rate.
Zaccanti et al.\ proposed this avalanche mechanism 
as an explanation for the loss features near the atom-dimer resonance. 
They also developed a model for calculating 
the number of atoms lost that demonstrated the plausibility 
of the avalanche mechanism \cite{Zaccanti:0904}.

In Ref.~\cite{LSB:1205}, we analyzed the avalanche mechanism for atom loss
and concluded that it was unable to produce a narrow loss feature 
near an atom-dimer resonance.
In this paper, we present a more thorough analysis of the 
avalanche mechanism.
We use Monte Carlo methods to generate avalanches of atoms 
that are initiated by recombination events
and then calculate the number of atoms lost
by averaging over avalanches.
We confirm that this number can be significantly 
larger than the naive value 3.
However, instead of a narrow peak in the atom loss rate 
near $a_*$, the avalanche mechanism produces a broad enhancement 
whose maximum is at a larger value of $a$.

This paper is organized as follows.
In Section~\ref{sec:Few-body}, we summarize the few-body physics 
that is used in our Monte Carlo model for the avalanche mechanism.
In Section~\ref{sec:Expinputs}, we describe the experimental inputs 
that are required in the Monte Carlo model.
In Section~\ref{sec:MonteCarlo}, we present the Monte Carlo method for
generating avalanches initiated by 3-body recombination events.
In Section~\ref{sec:Results}, we apply the Monte Carlo model 
to experiments on $^7$Li, $^{39}$K, and $^{133}$Cs atoms.
In Section~\ref{sec:Dimer-dimer}, we discuss the possibility
of enhanced atom losses near dimer-dimer resonances.

\section{Few-body physics}
\label{sec:Few-body}

In this section, we summarize the few-body physics that enters 
into our model for the avalanche mechanism.
An avalanche is initiated by
a 3-body recombination event in which three low-energy atoms collide 
to create an atom and a diatomic molecule,
which can be either the shallow dimer or a deep dimer.
The binding energy of the dimer is released in the kinetic energies 
of the outgoing atom and dimer.
In the case of the deep dimer, the outgoing atom and dimer have 
very high energies and therefore small cross sections, so
they escape from the trapping potential without any collisions.
In the case of the shallow dimer, the outgoing atom and dimer 
have large cross sections, so they may undergo secondary collisions.

\subsection{Few-body parameters}

We consider identical bosons of mass $m$ 
with a large positive scattering length $a$.
The universal few-body reaction rates associated with the 
zero-range limit are determined by three parameters \cite{Braaten:2004rn}:
\begin{itemize}
\item
the scattering length $a$, which can be controlled experimentally
by varying the magnetic field near a Feshbach resonance,
\item
the atom-dimer resonance $a_*$, or an equivalent Efimov parameter,
upon which physical observables can only depend log-periodically,
\item
a dimensionless parameter $\eta_*$, 
which controls the decay width of an Efimov trimer. 
\end{itemize}
The parameter $\eta_*$ is nonzero only if there are deep dimers 
that provide decay channels for the Efimov trimer.
The alkali atoms used in most cold atom experiments have many deep dimers.

In the negative-$a$ region, the most dramatic loss features are 
a sequence of narrow peaks in the 3-body recombination rate at the
3-atom resonances $(e^{\pi/s_0})^n a_{-}$,
where $e^{\pi/s_0} \approx 22.694$ is the discrete scaling factor, 
$n$ is an integer,
and $a_{-}$ differs from $a_{*}$ by a multiplicative constant:
\begin{equation}
a_{-} = -21.306~a_*.
\label{eq:a*-a*}
\end{equation}
The universal ratio $a_{-}/a_*$ was obtained with 5 digits of accuracy 
by dividing a 5-digit result for $a_{-} \kappa_*$ calculated by
Deltuva \cite{Deltuva:1202} by a 12-digit result 
for $a_* \kappa_*$ \cite{Braaten:2004rn}.
The parameters $a_{-}$ and $a_*$ related by 
Eq.~(\ref{eq:a*-a*}) are the scattering lengths at which the same 
Efimov trimer crosses the 3-atom and atom-dimer thresholds.
In the positive-$a$ region, the most dramatic loss features are 
a sequence of minima in the 3-body recombination rate at 
$(e^{\pi/s_0})^n a_{+}$, where $n$ is an integer
and $a_{+}$ differs from $a_{*}$ by a multiplicative constant:%
\footnote{Some papers follow Ref.~\cite{Grimm:0512} in using $a_+$
to denote $e^{-\pi/2 s_0}a_{+} = 0.93882~a_*$.  This is the position 
of a local maximum of $L_3/a^4$ but not of $L_3$,
where $L_3$ is the 3-body recombination rate constant.
The parameter $a_+$ is preferable as an Efimov parameter
because, in the limit $\eta_* \to 0$, 
it is the position of a zero of $L_3$ as well as $L_3/a^4$.
The resulting local minimum of $L_3$ is therefore a robust loss feature.} 
\begin{equation}
a_{+} = 4.4724~a_*.
\label{eq:a*0a*}
\end{equation}
The ratio $a_{+}/a_{-}= - e^{-\pi/2s_0}$ was obtained analytically 
by Hammer, Helfrich, and Petrov \cite{KHP:1001}.
The ratio $a_{+}/a_*$ in Eq.~(\ref{eq:a*0a*})
is obtained by multiplying this analytic result 
by that in Eq.~(\ref{eq:a*-a*}).
Either $a_{-}$ or $a_{+}$ can be used as the Efimov parameter 
in place of $a_*$. 
The ratios of the positions of loss features can differ 
from the universal ratios in Eqs.~(\ref{eq:a*-a*}) and (\ref{eq:a*0a*}) 
due to nonuniversal 
effects associated with a nonzero range.
Range corrections to the universal ratios have been analyzed by 
Ji, Phillips, and Platter \cite{Ji:2011qg}.

\subsection{Two-body observables}

The 2-body physics that enters into our model for the avalanche mechanism 
consists of the binding energy for the shallow dimer 
and the cross section for atom-atom scattering.
The universal binding energy for the shallow dimer is
\begin{equation}
E_d = \frac{\hbar^2}{m a^2}.
\label{eq:Edimer}
\end{equation}
The universal cross section for the elastic scattering of a pair of 
identical bosonic atoms with
center-of-mass wavenumber $k_{\rm cm}$ is
\begin{equation}
\sigma_{AA} = \frac{8 \pi a^2}{1 + a^2 k_{\rm cm}^2}.
\label{eq:sigmaAA}
\end{equation}
This universal expression is accurate if $k_{\rm cm}$ 
is much smaller than the inverse range. 

Three-body recombination at threshold creates an atom with wavenumber
$k = 2/(\sqrt{3} a)$.  The center-of-mass wavenumber
for its first collision is $k_{\rm cm} = 1/(\sqrt{3} a)$.
In the elastic collision of an energetic atom with a stationary atom, 
the kinetic energy of either outgoing atom is smaller than
that of the incoming atom by a factor whose average value is 1/2.
Thus the kinetic energies of the 
outgoing atoms decrease rapidly 
towards 0 as the avalanche develops.
The decreasing kinetic energies imply increasing atom-atom 
cross sections, although the increase is not dramatic.
If the recombination atom has many elastic collisions,
its cross section is larger than for its first collision 
by a factor of about 4/3.

\subsection{Three-body recombination rates}

\begin{figure}[t]
\vspace*{-0.0cm}
\includegraphics*[width=0.49\linewidth,angle=0,clip=true]{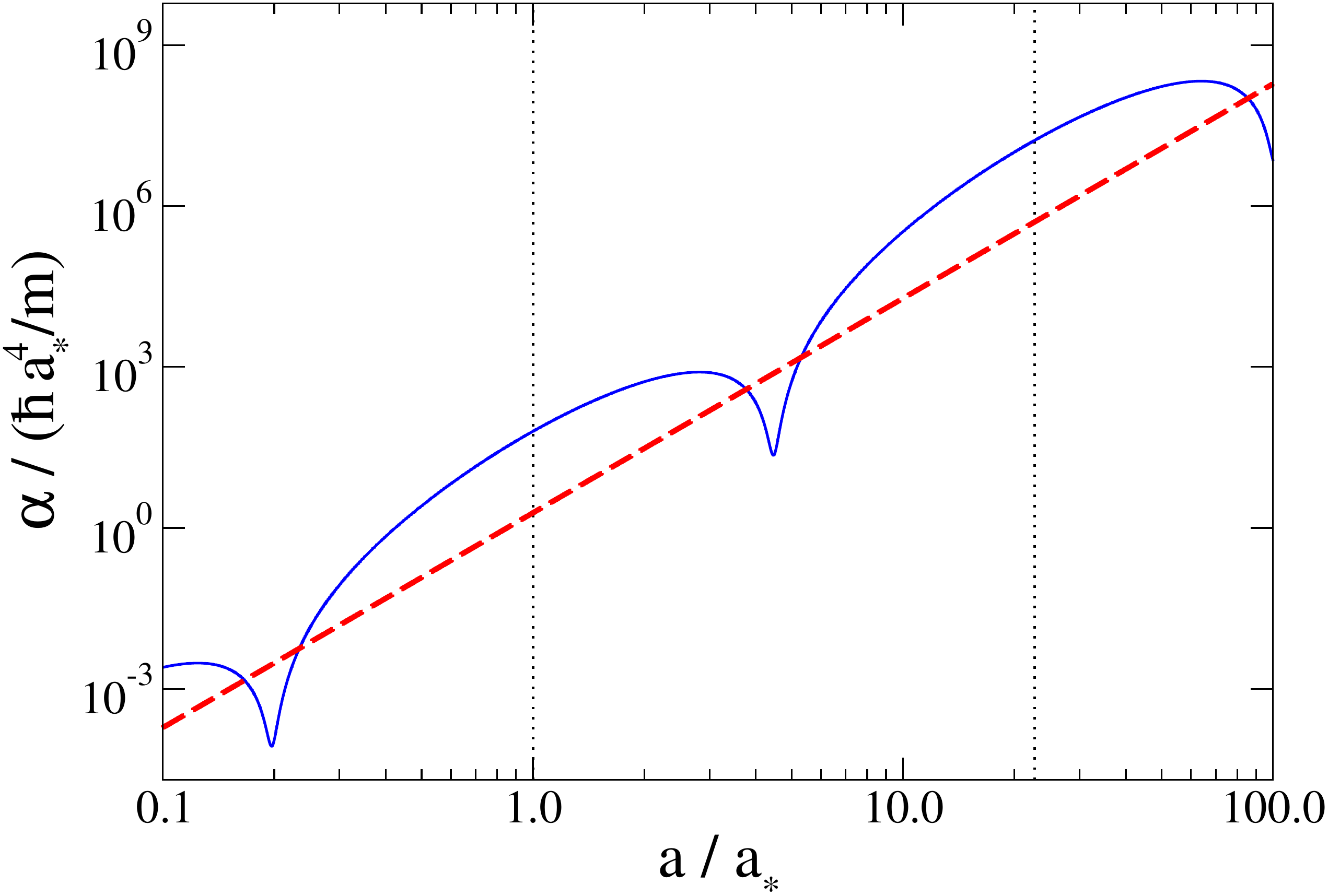}
\vspace*{0.0cm}
\caption{(Color online) 
Universal rate constants for 3-body recombination at threshold 
for $\eta_* = 0.03$ as functions of the scattering length.
The rate constant $\alpha_{\rm shallow}$ 
for recombination into the shallow dimer is shown as a solid (blue) line.
The rate constant $\alpha_{\rm deep}$
for recombination into deep dimers is shown as a dashed (red) line
that is almost straight.
The vertical dotted lines mark the positions of $a_*$ and $22.7~a_*$.
}
\label{fig:alpha}
\end{figure}

The 3-body recombination event that initiates an avalanche either
produces an atom and the shallow dimer or an atom and a deep dimer.
We consider systems of atoms in which the energy per particle 
is much smaller than the binding energy $E_d$ of the shallow dimer. 
We can therefore neglect the energies of the 
three incoming atoms and use the recombination rates at threshold.
If the system of atoms has number density $n$,
the recombination event rates can be expressed as $\alpha n^3$,
where $\alpha$ is a rate constant.
The universal event rate constants
$\alpha_{\rm shallow}$ and $\alpha_{\rm deep}$
for 3-body recombination at threshold into the shallow dimer 
and into deep dimers are conveniently expressed as functions of $a$,
$a_{+}$, and $\eta_*$ \cite{Braaten:2004rn}:
\begin{subequations}
\begin{eqnarray}
\alpha_{\rm shallow} &=&
\frac{128 \pi^2 (4 \pi - 3 \sqrt{3})
(\sin^2 [s_0 \ln (a/a_+)] + \sinh^2\eta_*)}
{\sinh^2(\pi s_0 + \eta_*) + \cos^2 [s_0 \ln (a/a_+)] }\,
\frac{\hbar a^4}{m} \,,
\label{alpha-shallow}
\\
\alpha_{\rm deep} &=&
\frac{128\pi^2(4\pi-3\sqrt{3}) \coth(\pi s_0) \cosh\eta_* \sinh\eta_*}
{\sinh^2(\pi s_0 + \eta_*) + \cos^2 [s_0 \ln (a/a_+)] }\,
\frac{\hbar a^4}{m} \,.
\label{alpha-deep}
\end{eqnarray} 
\label{alpha-shallowdeep}
\end{subequations}
These rate constants are shown as functions of $a$ in Fig.~\ref{fig:alpha} 
for $\eta_* = 0.03$.
Both rate constants have log-periodic modulation 
of $a^4$ scaling behavior.
For $\alpha_{\rm shallow}$, the log-periodic modulation produces 
local minima at scattering lengths near $(e^{\pi/s_0})^n a_{+}$
which become zeroes in the limit $\eta_* \to 0$.  
These minima arise from an Efimov interference effect.
For $\alpha_{\rm deep}$, the amplitude of the log-periodic modulation 
is too small to be evident in Fig.~\ref{fig:alpha}.  

\subsection{Atom-dimer cross sections}

The shallow dimer from 3-body recombination can collide 
with a low-energy atom in the cloud.
Since the energy per particle in the atom cloud is much smaller 
than $E_d$, we neglect the energy of the 
atom and use the atom-dimer cross sections for a stationary atom.
The collision between the shallow dimer and an atom can be
elastic, in which case the diatomic molecule 
in the final state is the shallow dimer, or inelastic, 
in which case it is a deep dimer.

\begin{figure}[t]
\vspace*{-0.0cm}
\includegraphics*[width=0.49\linewidth,angle=0,clip=true]{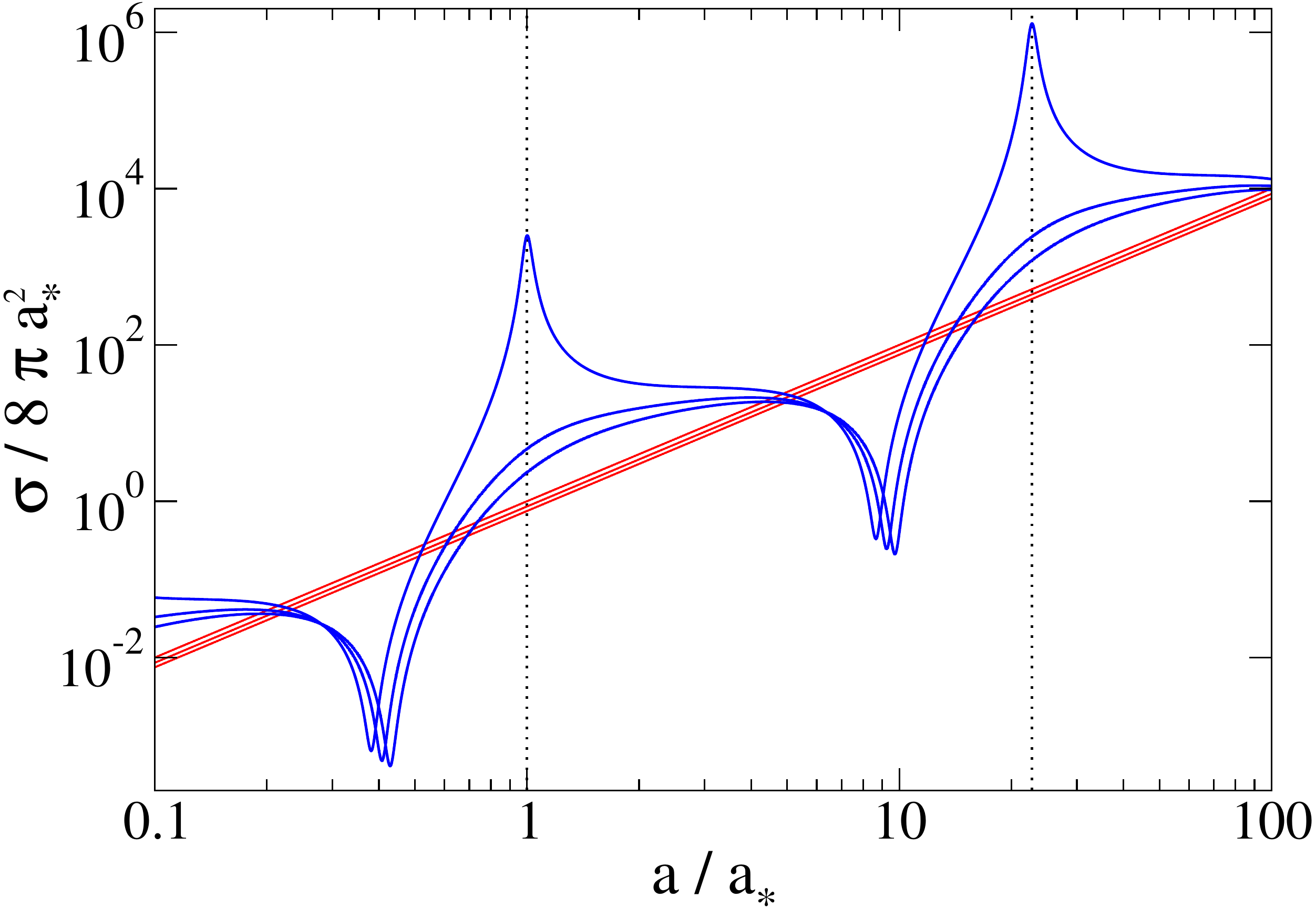}
\includegraphics*[width=0.49\linewidth,angle=0,clip=true]{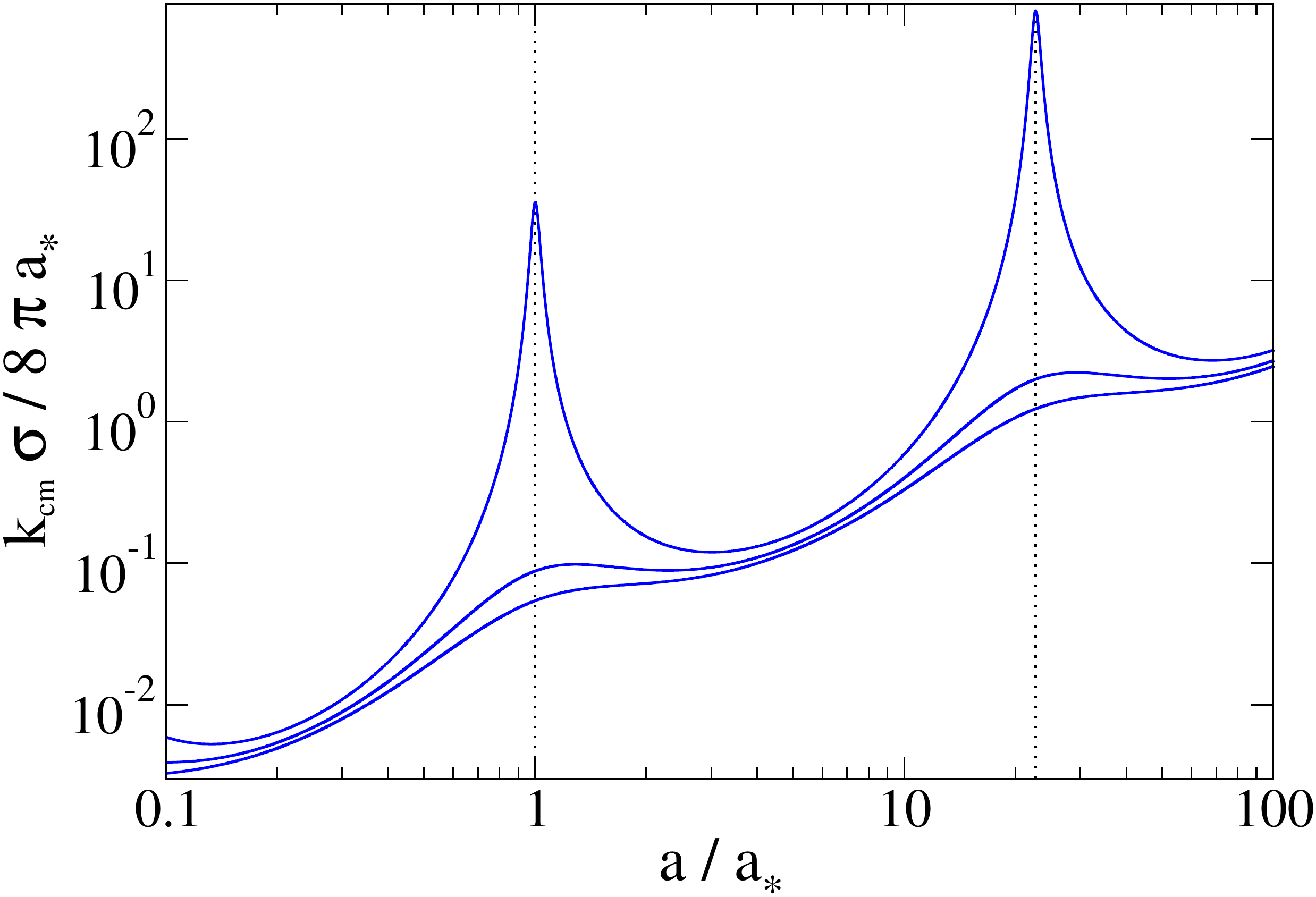}
\vspace*{0.0cm}
\caption{(Color online) 
Universal cross sections $\sigma_{AD}^{\rm (el)}$
for elastic atom-dimer scattering (left panel) and 
$k_{\rm cm}\sigma_{AD}^{\rm (in)}$ for inelastic 
atom-dimer scattering (right panel) for $\eta_* = 0.03$
as functions of the scattering length
for three different energies.
The vertical dotted lines mark the positions of $a_*$ and $22.7~a_*$.
The three curves
(in order of increasing cross sections at $a=a_*$)
are for the first scattering of the recombination dimer, 
a typical second scattering, and after many elastic scatterings.
In the left panel, the three straight closely-spaced (red) lines 
are the corresponding elastic cross sections for atom-atom collisions.  
}
\label{fig:sigma}
\end{figure}

The energy dependence of the atom-dimer cross sections is important.
In the collision of a dimer with wavenumber $k$
with a stationary atom, the
center-of-mass wavenumber $k_{\rm cm}$ is $k/3$.
An avalanche is initiated by a recombination event at threshold
creating an atom and a shallow dimer with wavenumbers
$k = 2/(\sqrt3 a)$.
In the first collision of the dimer with a stationary atom,
its center-of-mass wavenumber is $k_{\rm cm}=2/(3\sqrt3 a)$.
The collision energy in the center-of-mass frame is 
$E_{\rm cm} = \frac19 E_d$.
A subsequent elastic collision with a stationary atom decreases the
dimer's kinetic energy 
by a multiplicative factor whose average value is 5/9.
As the number of elastic collisions of the dimer increases,
$E_{\rm cm}$ decreases rapidly towards 0.

The universal elastic and inelastic atom-dimer cross sections 
$\sigma_{AD}^{\rm (el)}$ and $\sigma_{AD}^{\rm (in)}$
are conveniently expressed as functions of $a$, $a_*$, $\eta_*$, 
and the center-of-mass wavenumber $k_{\rm cm}$. 
The S-wave elastic atom-dimer cross section is given by
\begin{eqnarray}
\sigma_{AD}^{\rm (el)} = 
\frac{4 \pi}{\left| k_{\rm cm}\cot \delta_{AD}(k_{\rm cm})
-ik_{\rm cm}\right|^2} \,,
\label{sigma-ADel}
\end{eqnarray}
where $\delta_{AD}(k_{\rm cm})$ is the S-wave phase shift. 
The total S-wave cross section can be expressed via the optical theorem as
\begin{eqnarray}
\sigma_{AD}^{\rm (total)}  = \frac{4\pi}{k_{\rm cm}} \, {\rm Im} \,
\frac{1}{k_{\rm cm} \cot \delta_{AD} (k_{\rm cm}) -ik_{\rm cm}} \,.
\label{optical-AD}
\end{eqnarray}
The S-wave inelastic cross section is obtained by subtracting the 
elastic cross section from the total cross section:
\begin{eqnarray}
\sigma_{AD}^{\rm (in)} = \frac{4\pi}{k_{\rm cm}}
\frac{-{\rm Im}[ k_{\rm cm}\cot\delta_{AD}(k_{\rm cm})]}
    {\left| k_{\rm cm}\cot \delta_{AD}(k_{\rm cm})
-ik_{\rm cm}\right|^2} \,.
\label{sigma-ADin}
\end{eqnarray}
Efimov's radial law strongly constraints the dependence of the 
phase shift on $a_*$ \cite{Braaten:2004rn},
implying that it can be expressed as
\begin{eqnarray}
k a\,\cot \delta_{AD}(k) &=&
c_1(k a)
+ c_2(ka) \cot [s_0 \ln (a /a_*) + \phi(ka) + i \eta_*] \,.
\label{kacotdel:deep}
\end{eqnarray}
The functions $c_1(ka)$, $c_2(ka)$, and $\phi(ka)$ have been determined 
from the atom-dimer threshold $k=0$ to the dimer-breakup threshold
$k a = 2/\sqrt{3}$ by calculating the phase shifts
$\delta_{AD}(k)$ numerically using an effective field theory 
\cite{Braaten:2004rn}. The results
were parametrized as
\begin{subequations}
\begin{eqnarray}
c_1(ka)&=& -0.22 + 0.39 \, (k a)^2 -0.17 \, (k a)^4 \,,
\\
c_2(ka)&=& \phantom{+} 0.32 + 0.82 \, (k a)^2 -0.14 \, (k a)^4 \,,
\\
\phi(ka)&=& \phantom{+2.64}  - 0.83 \,(k a)^2 + 0.23 \, (k a)^4 \,.
\label{kcot-par:phi}
\end{eqnarray}
\end{subequations}

In the low-energy limit, the cross sections 
in Eqs.~(\ref{sigma-ADel}) and (\ref{sigma-ADin}) are determined by the 
atom-dimer scattering length $a_{AD}$, 
which can be expressed in the form \cite{Braaten:2004rn}
\begin{eqnarray}
a_{AD}  = 
\left( b_1 + b_2 \cot [s_0 \ln (a /a_*) + i \eta_*] \right) a \,,
\label{a-AD}
\end{eqnarray}
where $b_1$ and $b_2$ are universal numerical constants:
$b_1 \approx 1.46$, $b_2 \approx -2.15$.
In the limit $\eta_* \to 0$, 
$a_{AD}$ diverges at the atom-dimer resonance $a=a_*$.
The low-energy limit of the elastic cross section is
\begin{eqnarray}
\sigma_{AD}^{\rm (el)} \longrightarrow
4 \pi | a_{AD} |^2 \,.
\label{sigmaADel0}
\end{eqnarray}
The low-energy limit of the inelastic cross section 
multiplied by $k_{\rm cm}$ is
\begin{eqnarray}
k_{\rm cm} \sigma_{AD}^{\rm (in)} \longrightarrow 
4 \pi | a_{AD} |^2~{\rm Im}(1/a_{AD}) \,.
\label{sigmaADin0}
\end{eqnarray}
Both of the cross sections in Eqs.~(\ref{sigmaADel0}) and (\ref{sigmaADin0})
have a factor of $\sin^2[s_0 \ln (a /a_*)] + \sinh^2\eta_*$
in the denominator that produces a sharp peak 
at the atom-dimer resonance if $\eta_* \ll 1$.

The energy dependence of the universal atom-dimer cross sections 
is illustrated in Fig.~\ref{fig:sigma}.
The elastic cross section $\sigma_{AD}^{\rm (el)}$
and the inelastic cross section $k_{\rm cm}\sigma_{AD}^{\rm (in)}$
are shown as functions of $a$ for $\eta_* = 0.03$
at three different energies. 
These energies correspond to the first collision 
of the recombination dimer ($E_{\rm cm} = \frac19~E_d$), 
a typical second collision ($E_{\rm cm} = \frac{5}{81}~E_d$), 
and after many elastic collisions ($E_{\rm cm} \to 0$).
For the first collision, the elastic cross section 
has a broad peak as a function of $a$ with a local maximum near $4.34~a_*$,
which is close to the minimum
in the 3-body recombination rate:  $4.34~a_* \approx 0.97~a_+$.
The inelastic cross section 
for the first collision increases monotonically with $a$. 
For the typical second collision of the recombination dimer,
the elastic cross section is similar to that of the first collision,
but the inelastic cross section has a broad maximum just above $a_*$. 
After many elastic collisions,
both the elastic and inelastic atom-dimer cross sections peak sharply
near $a_*$.  

The atom-dimer cross sections in Eqs.~(\ref{sigma-ADel}) 
and (\ref{sigma-ADin}) are the S-wave contributions only.
There are also contributions from higher partial waves.
In the universal zero-range limit, the higher partial wave 
contributions to the inelastic cross sections vanish
and the higher partial wave contributions to the elastic cross sections 
are determined only by the scattering length $a$.
The contribution from the $L$'th partial wave
has the threshold behavior $(E_{\rm cm}/E_d)^L$.
The leading contribution is P-wave, and it is suppressed 
by a factor of $E_{\rm cm}/E_d$.  Thus its contribution to the atom-dimer 
elastic cross section for the first collision of the recombination dimer
($E_{\rm cm} = \frac19~E_d$) is expected to be about an order of magnitude 
smaller than the atom-atom cross section.
In Fig.~\ref{fig:sigma}, the P-wave atom-dimer cross section 
would be given by a straight line 
that is parallel to but significantly lower than the lowest 
straight line for the atom-atom cross section.
It would be completely negligible near an atom-dimer resonance at
$a=a_*$, but it would decrease the depths of the minima in the 
cross section.

\section{Experimental inputs}
\label{sec:Expinputs}

In this section, we summarize the experiments 
that have observed narrow loss features near an atom-dimer resonance. 
We identify the variables for these experiments
that are required as inputs to our Monte Carlo model for the avalanche
mechanism.

\subsection{Loss features near the atom-dimer resonance}

The first observation of a loss feature near an
atom-dimer resonance was by a group at Innsbruck in 2008 \cite{Grimm:0807}.
They used a thermal gas that was a mixture of $^{133}$Cs atoms 
and shallow dimers composed of those atoms.
They observed peaks in the atom and dimer loss rates at a scattering length 
near $+400~a_0$.  
The peak arises from the resonant enhancement of the inelastic scattering
of an atom and a shallow dimer into an atom and a deep dimer
due to an Efimov trimer near the atom-dimer threshold.
The Innsbruck group has also measured the loss rate for atom clouds 
consisting of $^{133}$Cs atoms only.  They did not observe 
any loss features near the atom-dimer resonance in systems
with atoms only.

There are three experiments that have observed loss features 
near an atom-dimer resonance in atom clouds that do not contain dimers.
The first such experiment was by a group at Florence in 2009 
using both a BEC and a thermal gas of $^{39}$K atoms \cite{Zaccanti:0904}.
They measured the 3-body loss rate constant $L_3$
as a function of the scattering length.  
They observed a peak in $L_3/a^4$ near $-1500~a_0$ that can be 
attributed to an Efimov trimer near the 3-atom threshold.  
They also observed two local minima in $L_3/a^4$ near 
$+224~a_0$ and $+5650~a_0$ that can be attributed to successive 
Efimov interference minima.  Of these three loss features, 
the highest precision in the determination of $\eta_*$ 
was obtained from the local minimum near $224~a_0$.  
We therefore use this loss feature to determine the Efimov parameters:
$a_{+} = 224~a_0$ and $\eta_{*} = 0.043$.
Given this value of $a_{+}$ and the universal ratio in Eq.~(\ref{eq:a*0a*}), 
atom-dimer resonances are predicted near
$50~a_0$ and $1140~a_0$.
The Florence group observed enhancements in the loss rate 
near $+30.4~a_0$ in a BEC and near $+930~a_0$ in a thermal gas,
both of which are reasonably close to the predicted atom-dimer resonances.
They attributed these loss peaks to the avalanche mechanism
of enhanced losses from secondary elastic collisions \cite{Zaccanti:0904}.

Another experiment in 2009 that observed a loss feature 
near an atom-dimer resonance was by a group at Rice University
using both a BEC and a thermal gas of $^{7}$Li atoms 
in the $|1,+1\rangle$ hyperfine state \cite{Hulet:0911}.
They observed two peaks in $L_3/a^4$
near $-298~a_0$ and near $-6301~a_0$ that can be attributed to
successive Efimov trimers near the 3-atom threshold.  
They also observed two local minima in $L_3/a^4$ near $+119~a_0$ 
and $+2676~a_0$ that can be attributed to successive 
Efimov interference features.  Of these four loss features, 
the highest precision in the determination of $\eta_*$ 
was obtained from the local minimum near $2676~a_0$.  
We therefore use this loss feature to determine the Efimov parameters:
$a_{+} = 2676~a_0$ and $\eta_{+} = 0.039$.
Given this value of $a_{+}$, an atom-dimer resonance is predicted at
$598~a_0$.
The Rice group observed an enhancement in $L_3/a^4$ in a BEC
near $+608~a_0$, which is close to the predicted atom-dimer resonance.

The third experiment that observed a loss feature 
near an atom-dimer resonance in an atom cloud that contained no dimers
was by a group at Bar-Ilan University
using a thermal gas of $^{7}$Li atoms in either the $|1,+1\rangle$ 
or $|1,0\rangle$ hyperfine state \cite{Khaykovich:1003}.
For both hyperfine states, they observed a peak in $L_3/a^4$
near $-270~a_0$ that can be attributed to an Efimov trimer near
the 3-atom threshold and a local minimum in $L_3/a^4$ near
$+1170~a_0$ that can be attributed to Efimov interference.  
A more thorough analysis was presented in Ref.~\cite{Khaykovich:1009}.
The Efimov parameters determined by fitting $L_3$
are $a_{+} = 1260~a_0$ and $\eta_* = 0.188$.
Given this value of $a_{+}$, an atom-dimer resonance is predicted at
$282~a_0$.  In Ref.~\cite{Khaykovich:1201},
additional data for $a$ below $220~a_0$ were presented,
revealing a narrow loss peak in $L_3$ near $+200~a_0$,
which is reasonably close to the predicted atom-dimer resonance.

\subsection{Experimental variables}
\label{sec:Expvar}

\begin{table}[t ]
 \begin{tabular}{|c||c|c|c|c|c|c|c|c|}
 \hline
 \multicolumn{1}{|c||}{}         &
 \multicolumn{2}{|c|}{$^7$Li}     &
 \multicolumn{2}{|c|}{$^{39}$K}   &
 \multicolumn{1}{|c|}{$^{133}$Cs} \\
 \hline
 \multicolumn{1}{|c||}{}         &
 \multicolumn{1}{|c|}{BEC \cite{Hulet:0911}}       &
 \multicolumn{1}{|c|}{thermal \cite{Khaykovich:1009}}   &
 \multicolumn{1}{|c|}{BEC \cite{Zaccanti:0904}}       &
 \multicolumn{1}{|c|}{thermal \cite{Zaccanti:0904}}   &
 \multicolumn{1}{|c|}{~thermal \cite{Grimm:1106}~~}   \\
 \hline
$\nu_x$ &
$236.0~{\rm Hz}$ & $1300.0~{\rm Hz}$ &
$75.0~{\rm Hz}$ & $75.0~{\rm Hz}$ &
$16.6~{\rm Hz}$ \\
$\nu_y$             &
$236.0~{\rm Hz}$ & $1300.0~{\rm Hz}$ &
$75.0~{\rm Hz}$  & $75.0~{\rm Hz}$ &
$18.31~{\rm Hz}$ \\
$\nu_z$ &
$4.6~{\rm Hz}$  & $190.0~{\rm Hz}$ &
$75.0~{\rm Hz}$ & $75.0~{\rm Hz}$ &
$3.78~{\rm Hz}$ \\
$N_0$ &
$4.0 \times 10^{5}$ & $3.5 \times 10^{4}$ &
$1.3 \times 10^{5}$ & $7.0 \times 10^{4}$ & 
$2.0 \times 10^{4}$ \\
$E_{\rm trap}$ &
$0.5 ~{\rm \mu K}$ & $7~{\rm \mu K}$    &
$1.0 ~{\rm \mu K}$ & $0.6 ~{\rm \mu K}$ & 
$0.2~{\rm \mu K}$ \\
$T$ & 
$<\!0.105 ~{\rm \mu K}$ & $1.4 ~{\rm \mu K}$ & 
$<\!0.17 ~{\rm \mu K}$  & $0.1 ~{\rm \mu K}$ & 
$0.015~{\rm \mu K}$ \\
$t_{\rm hold}$ & 
$\sim~0.003 ~{\rm s}$ & $0.01 - 5 ~{\rm s}$ & 
$1~{\rm s}$           & $0.17~{\rm s}$      & 
$\sim~1~{\rm s}$ \\
$a$ & 
~$500-4000~a_0$~  & ~$159-2663~a_0$~ & 
~$68-372~a_0$~ & ~$600-1548~a_0$~ & 
~$35-1270~a_0$~ \\
\hline
\end{tabular}
\caption{Experimental variables for experiments with 
$^{7}$Li, $^{39}$K, and $^{133}$Cs atoms: 
the trapping frequencies $\nu_x$, $\nu_y$, and $\nu_z$, 
the initial number of atoms $N_0$, 
the trap depth $E_{\rm trap}$, 
the temperature $T$, 
the holding time $t_{\rm hold}$,
and the range of scattering lengths $a$. 
In the case of a BEC, we give only an upper bound on the temperature $T$.
The upper bound is $\frac12 T_c$ for the BEC in 
Ref.~\cite{Hulet:0911} and $T_c$ for the  thermal gas in
Ref.~\cite{Zaccanti:0904}. 
}
\label{tab:Expvariables}
\end{table}

The important experimental variables in the measurements of
the loss rates of trapped atoms include the following:
\begin{itemize}
\item
the frequencies $\nu_x$, $\nu_y$, and $\nu_z$
of the harmonic trapping potential, which has the form
\begin{equation}
V(x,y,z) = 2 \pi^2 m
\left(  \nu_x^2 x^2 + \nu_y^2 y^2 + \nu_z^2 z^2 \right).
\label{Vxyz}
\end{equation}
\item
the initial number $N_0$ of trapped atoms.
\item
the temperature $T$ of the atoms.
\item
the trap depth $E_{\rm trap}$.
\item
the holding time $t_{\rm hold}$,
after which the remaining number $N$ of trapped atoms is measured.
\end{itemize}

In Table~\ref{tab:Expvariables},
we list the important experimental variables for five experiments
with $^7$Li atoms \cite{Hulet:0911,Khaykovich:1009},
$^{39}$K atoms \cite{Zaccanti:0904},
and $^{133}$ Cs atoms \cite{Grimm:1106}.
In two of the 5 experiments, the atom cloud was a BEC 
and in the other three, it was a thermal gas.
The experimental variables that are not given explicitly in the references
were obtained from private communications with the authors.
Different values of the experimental variables were used in 
different regions of the scattering length.
The values listed in Table~\ref{tab:Expvariables} are those that were used
in the range of $a$ given in the Table.

The holding time $t_{\rm hold}$ is generally chosen to be large enough 
that a significant fraction of the initial number $N_0$ of atoms are lost, 
so that this fraction can be measured with some precision.
The product of $t_{\rm hold}$ and a trapping frequency gives the number of
periods of the oscillation in that dimension before the atom number is
measured.  
The holding time is not used as an input in the Monte Carlo model
for simulating avalanches described in Section~\ref{sec:MonteCarlo}.

We use a simple model for the trap depth
that is specified by the single variable $E_{\rm trap}$.
Atoms and dimers that reach the edge of the atom cloud 
are assumed to be lost if their kinetic energies exceed 
$E_{\rm trap}$ and $2 E_{\rm trap}$, respectively.
Equivalently, this model for the trap depth can be expressed 
as a modification of the trapping potential 
for the atoms.  The potential for a single atom is given by 
Eq.~(\ref{Vxyz}) if $V(x,y,z) < E_{\rm trap}$ and is equal to 
the constant $E_{\rm trap}$ if $V(x,y,z) > E_{\rm trap}$.
The trap depth $E_{\rm trap}$ is generally substantially larger 
than the energy per atom $E/N$.  

The number of atoms lost also depends on the 
dimer binding energy $E_d = \hbar^2/ma^2$, which depends 
on the scattering length.
If $E_d < \frac32 E_{\rm trap}$, the recombination atom 
and the recombination dimer both remain trapped.
The dimer will eventually scatter inelasticly from an atom
in the cloud, so the number of lost atoms is 3.
If $ \frac32 E_{\rm trap} < E_d < 6 E_{\rm trap}$, 
the recombination dimer is trapped and must ultimately scatter inelasticly,
but the recombination atom is not trapped.
The largest possible number of lost atoms in the avalanche 
initiated by the recombination atom is the integer part of 
$(2/3) E_d/E_{\rm trap}$, which can be 
1, 2, or 3, depending on $a$.  
The dimer could also produce a single lost atom 
through an elastic collision, and it will eventually suffer an 
inelastic collision, resulting in the loss of 3 more atoms.  
Thus the maximum number of lost atoms increases from 3 to 7
as $E_d$ increases from $\frac32 E_{\rm trap}$ to $6 E_{\rm trap}$.
If $E_d > 6 E_{\rm trap}$, neither the atom nor the shallow dimer 
is trapped, so $N_{\rm lost}$ can be as large as $E_d/E_{\rm trap}+3$.
Our simple model for the trap depth
implies discontinuities in physical observables at 
$E_d = \frac32 E_{\rm trap}$ and $E_d = 6 E_{\rm trap}$. 
Thus a more elaborate model is probably required to give 
accurate predictions for the number of lost atoms in
a region that includes the interval
$ \frac32 E_{\rm trap} < E_d < 6 E_{\rm trap}$.

\subsection{Number densities}

The frequencies $\nu_x$, $\nu_y$, and $\nu_z$,
the initial number $N_0$ of trapped atoms,
and the temperature $T$ determine the number density $n(x,y,z)$ of the atoms.
We consider three simple cases for the system of trapped atoms:
\begin{itemize}
\item
a Bose-Einstein condensate (BEC) of atoms at 0 temperature 
in the Thomas-Fermi limit,
\item
a thermal gas of atoms in the weak-interaction limit 
at the critical temperature $T_c$,
\item
a thermal gas of atoms in the weak-interaction limit 
at a temperature $T$ much larger than $T_c$.
\end{itemize}

In a BEC of atoms at zero 
temperature in the Thomas-Fermi limit, the number density 
depends on the scattering length $a$:
\begin{equation}
n(x,y,z) = \frac{m}{4 \pi \hbar^2 a}
{\rm max}\{ \mu(a) - V(x,y,z), 0 \},
\label{nBEC}
\end{equation}
where $\mu(a)$ is the chemical potential, which also depends on $a$:
\begin{equation}
\mu(a) = \frac{\hbar^2}{2 m}
\left( \frac{15 N a}{a_x^2 a_y^2 a_z^2} \right)^{2/5}.
\label{muTF}
\end{equation}
The trap lengths $a_x$, $a_y$, and $a_z$ are determined by the 
trapping frequencies: $a_i = (\hbar/2 \pi m \nu_i)^{1/2}$.
The energy per atom is just the chemical potential: 
\begin{equation}
E/N = \mu(a).
\end{equation}

The critical temperature for Bose-Einstein condensation 
in the trapping potential is 
\begin{equation}
k_B T_c = \frac{\hbar^2}{m}
\left( \frac{N}{\zeta(3)\, a_x^2 a_y^2 a_z^2} \right)^{1/3},
\label{k_B Tc}
\end{equation}
where $\zeta(3) \approx 1.202$.
For a thermal cloud of trapped atoms above $T_c$, 
the appropriate phase-space distribution in the weak-interaction limit 
is the Bose-Einstein distribution.
At the critical temperature, the number density
can be expressed in terms of a polylogarithm:
\begin{equation}
n(x,y,z) = 
\left( \frac{m k_B T_c}{2 \pi \hbar^2} \right)^{3/2}
{\rm Li}_{3/2}(\exp(- V(x,y,z)/k_B T_c)).
\label{ncritical}
\end{equation}
The energy per atom at $T_c$ is
\begin{equation}
E/N = \frac{\pi^4}{30 \zeta(3)} k_B T_c,
\label{Ec}
\end{equation}
which is approximately $2.7012~k_B T_c$.

If $T$ is large enough compared to $T_c$, the Bose-Einstein distribution 
can be approximated by the Boltzmann distribution.
The number density then reduces to a Gaussian:
\begin{equation}
n(x,y,z) = 
\frac{N \lambda_T^3}{8 \pi^3 a_x^2 a_y^2 a_z^2}
\exp(- V(x,y,z)/k_B T),
\label{nBoltzmann}
\end{equation}
where $\lambda_T = (2 \pi \hbar^2/m k_B T)^{1/2}$ is the 
thermal quantum wavelength.  The energy per atom is given 
by the equipartition theorem: 
\begin{equation}
E/N = 3 k_B T.
\end{equation}

\subsection{Loss rate and heating rate}

The rate at which the local number density $n(\bm{r})$ of atoms 
in a thermal gas decreases due to 3-body recombination 
can be expressed as a local differential equation:
\begin{equation}
\frac{d\ }{dt} n(\bm{r}) = 
- \left( N_{\rm lost}(\bm{r})  \alpha_{\rm shallow} 
        + 3 \alpha_{\rm deep} \right) n^3(\bm{r}),
\label{dndt}
\end{equation}
where $N_{\rm lost}(\bm{r})$ is the average number of atoms lost 
from a recombination event that creates a shallow dimer at the point $\bm{r}$.
Upon integrating Eq.~(\ref{dndt}) over space, we obtain the rate 
at which the number $N$ of atoms decreases:
\begin{equation}
\frac{dN}{dt}  = 
- \left( \langle N_{\rm lost} \rangle \alpha_{\rm shallow} 
        + 3 \alpha_{\rm deep} \right) \langle n^2 \rangle N,
\label{dNdt}
\end{equation}
where $\langle n^2 \rangle$ and $\langle N_{\rm lost} \rangle$
are spacial averages weighted by $n(\bm{r})$ and $n^3(\bm{r})$,
respectively. Equivalently, $\langle N_{\rm lost} \rangle$ 
is the number of atoms lost in a single avalanche 
averaged over the probability distribution for avalanches.
In the case of a BEC, the right sides of Eqs.~(\ref{dndt}) and (\ref{dNdt})
should be multiplied by 1/6 to take into account that the 3 atoms 
undergoing recombination are identical bosons.
The loss rate constant $L_3$ is defined to be the coefficient of 
$-\langle n^2 \rangle N$ in $dN/dt$ for thermal gas:
\begin{equation}
L_3 = \langle N_{\rm lost} \rangle \alpha_{\rm shallow} 
        + 3 \alpha_{\rm deep} .
\label{L3-alpha}
\end{equation}
Measurements of $L_3$ in a BEC and in a thermal gas of the same atoms
should agree to within experimental uncertainties.

Atoms in the avalanche with kinetic energy less than $E_{\rm trap}$
can never escape from the trapping potential.
Through subsequent elastic collisions, their kinetic energy 
is ultimately transformed into heat.
If most of the heat is deposited near the recombination point,
the rate at which a thermal gas of trapped atoms gains heat $Q$ 
from the avalanche mechanism is 
\begin{equation}
\frac{dQ}{dt}  = 
\langle E_{\rm heat} \rangle \alpha_{\rm shallow} \langle n^2 \rangle N,
\label{dQdt}
\end{equation}
where $\langle E_{\rm heat} \rangle$ 
is the average amount of energy transformed into heat 
in a single avalanche.
In the case of a BEC, the right side of Eqs.~(\ref{dQdt})
should be multiplied by 1/6 to take into account that the 3 atoms 
undergoing recombination are identical bosons.

The number density profiles in Eq.~(\ref{nBEC}) for a BEC 
and in Eq.~(\ref{nBoltzmann}) for a thermal gas 
are those that would be expected in the absence 
of atom loss processes, such as 3-body recombination.
Loss processes decrease the number $N$ of atoms,
allow energy to be carried out of the system by the lost atoms,
and also add heat $Q$ to the system.  
These effects can change the number density
and the energy density of atoms as functions of time.
In the case of a BEC, the atom loss and heating can generate 
a thermal cloud inside and surrounding the BEC.
If too much energy is added to the system,
its temperature can be raised above the critical temperature
for Bose-Einstein condensation, in which case the BEC component 
disappears completely.

In the case of a thermal gas, the atom loss and the heating 
change the number of atoms $N$ and their total energy.
If the thermalization rate is sufficiently fast, 
the number density can still be approximated
by the density profile in Eq.~(\ref{nBoltzmann})
with time-dependent $N$ and $T$.
To measure the loss rate constant $L_3$, 
that time dependence must be taken into account.
A method for doing this was developed in Ref.~\cite{Grimm:0304}.
The coupled rate equations for $N$ and $T$
(in the absence of background gas collisions) were 
expressed in the form
\begin{subequations}
\begin{eqnarray}
\frac{dN}{dt} &=& - \frac{\gamma N^3}{T^3},
\label{dNdt:Grimm}
\\
\frac{dT}{dt} &=& 
\frac{\gamma (T+T_h) N^2}{3 T^3},
\label{dTdt:Grimm}
\end{eqnarray}
\label{dNTdt:Grimm}
\end{subequations}
where $\gamma$ and $T_h$ are constants.
The solutions to these coupled differential equations 
depend on $\gamma$ and $T_h$.
If $N(t)$ is measured as a function of the holding time $t$,
the two parameters can be adjusted to fit that time dependence.
The rate constant $L_3$ can then be determined from the fitted value 
of $\gamma$: 
\begin{equation}
L_3 = \left( \frac{\sqrt{3} k}{2 \pi m \bar \nu^2 } \right)^3 \gamma ,
\label{L3-gamma}
\end{equation}
where $\bar \nu = (\nu_x \nu_y \nu_z)^{1/3}$ is the geometric mean of the
trapping frequencies.

We can derive the coupled equations for $N$ and $T$ in 
Eqs.~(\ref{dNTdt:Grimm}) from our rate equations for $N$ and $Q$
in Eqs.~(\ref{dNdt}) and (\ref{dQdt}).
This derivation determines the fitting parameters $\gamma$ and $T_h$ 
in terms of the quantities
$\langle N_{\rm lost} \rangle$ and $\langle E_{\rm heat} \rangle$
associated with the avalanche mechanism.
The total energy $E$ of the thermal gas in a harmonic trap
is $E = 3 N k_B T$.
It changes because the recombination event delivers energy
to an atom and a dimer,
the lost atoms carry away their kinetic energy,
and the atoms that are elasticly scattered 
but remain trapped deposit their energy as heat.
The average energy of an atom in the thermal cloud
is $3 k_B T$.
Since the recombination probability is proportional to $n^3(x,y,z)$,
the incoming atoms in a 3-body recombination event have a smaller
average energy $2k_B T$.  If all the lost atoms originate near the 
recombination point, their average energy is also $2k_B T$.
Thus the rate of change in the total energy is 
\begin{equation}
\frac{dE}{dt} = 
 - \left( \langle N_{\rm lost} \rangle \alpha_{\rm shallow} + 3 \alpha_{\rm deep} \right) 
 \langle n^2 \rangle N (2 k_B T) 
 + \frac{dQ}{dt}.
\label{dEdt}
\end{equation}
Setting $E=3Nk_B T$ and using Eqs.~(\ref{dNdt}) and (\ref{dQdt}) for
$dN/dt$ and $dQ/dt$, we can obtain a rate equation for $T$:
\begin{eqnarray}
\frac{dT}{dt} = 
\left( \frac{\alpha_{\rm shallow} \langle E_{\rm heat} \rangle}{3 k_B T}
+ \frac{\langle N_{\rm lost} \rangle \alpha_{\rm shallow} 
       + 3 \alpha_{\rm deep}}{3} 
\right) \langle n^2 \rangle T.
\label{dTdt}
\end{eqnarray}
The density-weighted average $\langle n^2 \rangle$
in a thermal gas is
\begin{equation}
\langle n^2 \rangle = \frac{1}{3\sqrt{3}}
\left( \frac{N \lambda_T^3}{8 \pi^3 a_x^2 a_y^2 a_z^2} \right)^2 .
\label{<n^2>}
\end{equation}
Since this is proportional to $N^2/T^3$, 
the rate equations for $N$ in Eq.~(\ref{dNdt}) 
and $T$ in Eq.~(\ref{dTdt}) do have the form given in
Eqs.~(\ref{dNTdt:Grimm}).  The constant $\gamma$ is 
proportional to the rate constant $L_3$ in Eq.~(\ref{L3-alpha})
in accord with Eq.~(\ref{L3-alpha}).
The product of the constants $\gamma$ and $T_h$ in Eq.~(\ref{dTdt:Grimm}) 
is determined by $\langle E_{\rm heat} \rangle$ only:
\begin{equation}
\gamma T_h = \left( \frac{2 \pi m \bar \nu^2}{\sqrt{3} k} \right)^3 
\frac{\alpha_{\rm shallow} \langle E_{\rm heat} \rangle}{k}.
\label{gammaTh}
\end{equation}
In Ref.~\cite{Grimm:0304}, $k_B T_h$ was interpreted 
as the energy per lost atom.
Combining Eqs.~(\ref{gammaTh}) and (\ref{L3-gamma}) 
with the expression for $L_3$ in Eq.~(\ref{L3-alpha}),
we see that $k_B T_h$ is indeed equal to $E_{\rm heat}$
if $\alpha_{\rm deep}$ is negligible compared to $\alpha_{\rm shallow}$.

In Section~\ref{sec:MonteCarlo}, we develop a Monte Carlo model  
for the avalanche mechanism that can be used to 
calculate $N_{\rm lost}$ and $E_{\rm heat}$.
These quantities can also be determined experimentally
using the values of $\gamma$ and $T_h$ obtained by fitting the 
time dependence of $N(t)$.
By comparing the calculated and measured values of 
$N_{\rm lost}$ and $E_{\rm heat}$,
we could test our Monte Carlo model for the avalanche mechanism 
and perhaps develop a more accurate description of the loss process.

\section{Monte Carlo method}
\label{sec:MonteCarlo}

In this section, we describe our Monte Carlo model for the avalanche
mechanism.  
We also compare it to previous models for the avalanche mechanism.

\subsection{Approximations}
\label{sec:approx}

The important energy scales in cold atom experiments
include the energy per atom $E/N$,       
the trap depth $E_{\rm trap}$,
and the dimer binding energy $E_d=\hbar^2/ma^2$,
which depends on the scattering length $a$.
For a thermal gas with temperature $T$, $E/N$ is $3 k_B T$. 
For a BEC, $E/N$ is equal to the chemical potential $\mu(a)$
given in Eq.~(\ref{muTF}), which depends on $a$.  
Another relevant energy scale is $2.7~k_B T_c$, 
which is the energy per atom at the critical temperature.
In the case of a thermal gas, $E/N$ must be significantly larger 
than $2.7~k_B T_c$ in order to use the Boltzmann distribution
instead of the Bose-Einstein distribution.
In the case of a BEC, $N(2.7~k_B T_c)$ is roughly the heat energy 
that must be added to change it to a thermal gas. 
The various energy scales are listed in Table~\ref{tab:Escales}
for each of the five sets of experimental variables listed in
Table~\ref{tab:Expvariables}. 

Our simple model for the trap depth is described in 
Section~\ref{sec:Expvar}.
If an energetic atom or dimer reaches the edge of the cloud, 
it is lost from the trap if its kinetic energy is greater than 
$E_{\rm trap}$ or $2 E_{\rm trap}$, respectively.
Otherwise it will follow a curved trajectory that returns to the cloud. 
A trapped atom that returns to the cloud will eventually thermalize 
through elastic collisions, transforming its kinetic energy into heat.
A trapped dimer that returns to the atom cloud 
will eventually suffer an inelastic collision that results 
in the loss of 3 atoms.  Before the inelastic collision,
it could scatter elasticly,
transforming some of its kinetic energy into heat,
but we ignore that small contribution to the heat.

The trap depth $E_{\rm trap}$ is usually 
substantially larger than the energy per particle $E/N$.  
Otherwise, atoms will be rapidly lost from the trap 
until most of the atoms have energy smaller than $E_{\rm trap}$.  
The energy per particle is more 
than an order of magnitude smaller than $E_{\rm trap}$ 
over most of the range of scattering length for most of 
the experiments listed in Table~\ref{tab:Expvariables}.
The exceptions are the $^7$Li BEC experiment at the upper end of 
the range of $a$, where $E/N$ is about $0.5~E_{\rm trap}$
and the $^7$Li and $^{39}$K thermal gas experiments, 
in which $E/N$ is also about $0.5~E_{\rm trap}$.  In these cases, 
our simple model for the trap depth may not be 
sufficient to calculate the effects of the avalanche mechanism 
accurately.

As $a$ is increased by adjusting the magnetic field, 
the dimer binding energy $E_d$ can decrease from much larger 
than $E_{\rm trap}$ to much smaller than $E_{\rm trap}$.
However it usually remains much larger than $E/N$.  
This allows the kinetic energies of the atoms in the cloud to be ignored in 
few-body reaction rates.  
For the experiments listed in Table~\ref{tab:Expvariables}, $E_d$ is 
more than an order of magnitude 
larger than $E/N$ over most of the range of scattering lengths.  
There are a few exceptions near the upper ends of the ranges of $a$.
In the $^7$Li BEC experiment and the $^{39}$K thermal gas experiment,
$E_d$ becomes as small as $6~E/N$ at the largest values of $a$.
In the $^7$Li thermal gas experiment, 
$E_d$ becomes as small as $0.8~E/N$ at the largest value of $a$.

 \begin{table}[t ]
 \begin{tabular}{|c||c|c|c|c|c|c|c|c|}
 \hline
 \multicolumn{1}{|c||}{}         &
 \multicolumn{2}{|c|}{$^7$Li}     &
 \multicolumn{2}{|c|}{$^{39}$K}   &
 \multicolumn{1}{|c|}{$^{133}$Cs} \\
 \hline
 \multicolumn{1}{|c||}{}         &
 \multicolumn{1}{|c|}{BEC \cite{Hulet:0911}}       &
 \multicolumn{1}{|c|}{~thermal \cite{Khaykovich:1009}~~}   &
 \multicolumn{1}{|c|}{BEC \cite{Zaccanti:0904}}       &
 \multicolumn{1}{|c|}{~thermal \cite{Zaccanti:0904}~~}   &
 \multicolumn{1}{|c|}{~thermal \cite{Grimm:1106}~~}   \\
 \hline
$E/N$ &
$~0.10-0.23~$ & $~4.2$     & $~0.049-0.10~$ & $~0.3$    & $~0.045$ \\
$~2.7~k_B T_c~$ & 
$0.57$         & $~2.7$     & $~0.46$        & $~0.38$   & $~0.035$ \\
$E_{\rm trap}$ & 
$0.5$          & $~7$       & $~1.0$         & $~0.6$    & $~0.2$   \\
$E_d$ & 
$100-1.5$     & $~980-3.5$ & $~960-32$      & $~12-1.9$ & $~1040-0.81$ \\
\hline
\end{tabular}
\caption{Energy scales in $\mu K$ for experiments with
$^{7}$Li, $^{39}$K, and $^{133}$Cs atoms:    
the energy per atom $E/N$,       
the energy per atom at the critical temperature $2.7~k_B T_c$,
the trap depth $E_{\rm trap}$,
and the range of the dimer binding energy $E_d=\hbar^2/ma^2$.
The ranges of $E/N$ for a BEC and the ranges of $E_d$
correspond to the ranges of $a$ given in Table~\ref{tab:Expvariables}.
}
\label{tab:Escales}
\end{table}

If the atom cloud is a BEC in the Thomas-Fermi limit,
the trajectory of an energetic atom inside the BEC is a straight line,
because the trapping potential energy of an atom and its 
mean-field energy add up to the constant chemical potential $\mu(a)$.
If the atom flies beyond the edge of the BEC, it follows a curved 
trajectory determined by the harmonic potential.
The trajectory of an energetic dimer is curved even inside the BEC,
because its mean-field energy differs from that of a pair of atoms.
If the atoms are in a thermal cloud,
the trajectory of an energetic atom or energetic dimer
is always curved.
However if the kinetic energy of the atom or dimer is large enough that 
it can transfer an energy greater than $E_{\rm trap}$
to an atom in the cloud through an elastic collision,
its trajectory has small curvature, 
and it can be approximated by a straight line.
The kinetic energy of an atom or the dimer can change between 
scattering points,
because the potential energy (and, in the case of a BEC, 
the mean-field energy) depends on the position in the cloud.
However if the kinetic energy of the atom or dimer is large enough that 
it can transfer an energy greater than $E_{\rm trap}$
to an atom in the cloud, the change in the kinetic energy is negligible.

The rate equations for $N$ and $Q$ in Eqs.~(\ref{dNdt}) 
and (\ref{dQdt}) were derived 
from the local rate equation for $n(\bm{r})$ in Eq.~(\ref{dndt}).  
However the avalanche mechanism 
makes the loss process partly nonlocal.
Some of the atoms that escape from the trapping potential
receive their kinetic energy from an elastic collision 
at a scattering point $(x,y,z)$ that may not be near the 
recombination point $(x_0,y_0,z_0)$.  
The distance $[(x-x_0)^2+(y-y_0)^2+(z-z_0)^2]^{1/2}$
is not a good measure of the nonlocality,
because the length scales set by the trapping potential 
are different in different directions.
A better measure of the nonlocality is the dimensionless distance
\begin{equation}
{\hat\ell}^{\,2} = 
\frac{(x-x_0)^2}{a_x^2} + \frac{(y-y_0)^2}{a_y^2}
+ \frac{(z-z_0)^2}{a_z^2}.
\label{l-hat}
\end{equation}
which is the square of the number of oscillator lengths 
separating the recombination point 
and the scattering point.
The local approximation for the loss rate 
will be valid if $\langle {\hat\ell}^{\,2} \rangle \ll 1$,
where the average is over lost atoms and over avalanches.
If there are no collisions, the scattering point coincides 
with the recombination point and ${\hat\ell}^{\,2} = 0$.
In general, $\langle {\hat\ell}^{\,2} \rangle$ depends on the prescription 
for the average.  Since the atoms are identical bosons,
a lost atom could be identified with any of the stationary atoms in the 
chain of previous elastic scatterings or with one of the three 
incoming atoms in the recombination event.
Thus the scattering point $(x,y,z)$ for a lost atom
can be taken as its point of last scattering or the recombination point 
or any of the collision points in between.
The atoms composing a dimer that escapes or scatters inelasticly 
could be identified with two of the incoming atoms in the recombination event,
but they also could be identified with any of the stationary atoms 
from which the dimer scattered elasticly.
One possible prescription is to choose the scattering point $(x,y,z)$
to be the first collision point for any of its ancestors
in the binary tree with equal probability.
A more reasonable prescription is to choose the scattering point $(x,y,z)$
to be the collision point at which the greatest energy is 
imparted to the lost atom or to one of its ancestors.
With this prescription, the 3 lost atoms 
from an inelastic atom-dimer collision will be 
assumed to come from the inelastic collision point.
For those atoms that are lost individually, the most energetic will be 
assumed to come from the recombination point with ${\hat\ell}^{\,2} = 0$.
The other lost atoms will usually be assumed to come from  
one of the first elastic collisions after the recombination event.
This prescription is likely to give
$\langle {\hat\ell}^{\,2} \rangle \ll 1$, thus providing
some justification for the local approximation.

We now list the most important approximations made 
in our Monte Carlo model:
\begin{itemize}
\item
~We neglect the energies of the low-energy atoms in the atom cloud.
\item
We approximate the trajectories of the dimer 
and the atoms between scattering events by straight lines.
\item
We take the momentum of an incoming atom or dimer in a collision
to be the same as that particle's outgoing momentum from the previous
scattering event.
\item
When comparing the energy of an atom or dimer to the trap depth,
we ignore its potential energy (and, 
in the case of a BEC, its mean-field energy). 
\item
We make the local approximation  
that most of the lost atoms come from near the recombination point
and also that most of the heat from scattered atoms 
that are not lost is deposited near the recombination point.
\end{itemize}

\subsection{Simulating avalanches}
\label{sec:Simulation}

The development of an avalanche can be decomposed into discrete steps
corresponding to the recombination event and the subsequent scattering events.
Given the state of the avalanche immediately before each event, 
the state immediately after the event has a simple probability distribution.  
All these simple probability distributions together determine 
the probability distribution of avalanches.
We can generate avalanches with this probability distribution 
using a Monte Carlo method.  
At each of the events in the evolution of the avalanche, we 
use a random number generator to determine the subsequent state.
The simple probability distributions can be generated as follows:
\begin{itemize}
\item
The position $(x,y,z)$ of the recombination point,
whose probability distribution is proportional to $n^3(x,y,z)$,
is determined by three random numbers.
\item
The outgoing wavevectors $\bm{k}$ and $\bm{k}'$
for a pair of scattered particles
are determined by the incoming wavevectors and two random numbers.
In the center-of-momentum frame, the distribution 
of the wavevectors $\pm {\bm k}_{\rm cm}$ is isotropic.
\item
Whether or not an atom or dimer produced by the recombination event 
or a scattering event is scattered before it reaches the edge of the 
atom cloud is determined by whether the scattering probability 
$1- \exp(- \sigma \int \!n\, \hbox{d}\ell)$
is greater than or less than a random number between 0 and 1.
The cross section $\sigma$ is $\sigma_{AA}$ if the particle is an atom
and $\sigma_{AD}^{\rm (el)} + \sigma_{AD}^{\rm (in)}$  if it is a dimer.
The column density $\int \!n\, \hbox{d}\ell$ is calculated by integrating 
from the position of the recombination or scattering event
out to infinity along a straight line
in the direction of the wavevector $\bm{k}$ of the particle.
If the atom or dimer scatters,
the same random number is used to determine the position 
of its scattering event by solving for the length $\ell$ 
along the path for which $1 - \exp(- \sigma \int_0^\ell \!n\, \hbox{d}\ell)$
is equal to the random number.
\item
Given that a dimer scatters, it scatters inelasticly 
if the probability 
$\sigma_{AD}^{\rm (in)}/(\sigma_{AD}^{\rm (el)} + \sigma_{AD}^{\rm (in)})$
is greater than a random number between 0 and 1.
Otherwise, the dimer scatters elasticly.
\end{itemize}

\subsection{Atom loss and heating}

The Monte Carlo method described in Section~\ref{sec:Simulation} 
generates a binary tree.
The initial node, which represents the recombination event,
has two branches corresponding to the dimer and the atom.
For every elastic scattering event, there is a node with two branches
that correspond to the two outgoing particles.
Finally there are terminal nodes associated with atoms or dimers 
whose ultimate fate has been determined.
More specifically, the terminal nodes corrrespond to
atoms or dimers that are lost,
atoms or dimers that are trapped, 
and dimers that have inelastic collisions.
Each terminal node gives a contribution 
$\Delta N_{\rm lost}$ to the number of atoms lost
and $\Delta E_{\rm heat}$ to the heat of the remaining atoms.
The conditions for a branch to end at a terminal node 
and the corresponding values of $\Delta N_{\rm lost}$ and 
$\Delta E_{\rm heat}$ are as follows:
\begin{itemize}
\item
If an outgoing atom from a scattering event 
has kinetic energy $E < E_{\rm trap}$,
it remains trapped:
$\Delta N_{\rm lost} = 0$ and $\Delta E_{\rm heat} = E$.
\item
If an atom that reaches the edge of the atom cloud
has kinetic energy $E > E_{\rm trap}$, it is lost:
$\Delta N_{\rm lost} = 1$ and $\Delta E_{\rm heat} = 0$.
\item
If a dimer has an inelastic collision, 
both it and the scattered atom are lost:
$\Delta N_{\rm lost} = 3$ and $\Delta E_{\rm heat} = 0$.
\item
If a dimer that reaches the edge of the atom cloud
has kinetic energy $E > 2 E_{\rm trap}$, it is lost:
$\Delta N_{\rm lost} = 2$ and $\Delta E_{\rm heat} = 0$.
\item
If a dimer that reaches the edge of the atom cloud
has kinetic energy $E < 2 E_{\rm trap}$,
it will return to the cloud 
and will eventually suffer an inelastic collision:  
$\Delta N_{\rm lost} = 3$ and $\Delta E_{\rm heat} = 0$.
We ignore any heat from additional elastic collisions
before the final inelastic collision.
\end{itemize}
The number of terminal nodes in the binary tree is 2
if the cross section and the column density are small enough 
that there is no scattering.  The number of terminal nodes 
is generally larger in a BEC than in a thermal gas.
For the sets of experimental variables listed in
Table~\ref{tab:Expvariables},
the number of terminal nodes is sometimes greater than 50.

The quantities $N_{\rm lost}$ and 
$E_{\rm heat}$ for a single avalanche 
are obtained by adding up $\Delta N_{\rm lost}$ 
and $\Delta E_{\rm heat}$ for all the terminal nodes.
Their averages $\langle N_{\rm lost} \rangle$ and 
$\langle E_{\rm heat} \rangle$ are calculated by averaging over many 
avalanches generated using the Monte Carlo method described 
in Section~\ref{sec:Simulation}.
These averages have discontinuities as functions of $a$ at 
$E_d = \frac32 E_{\rm trap}$ and $E_d = 6 E_{\rm trap}$,
which are artifacts of our simple model for 
the trap depth.  Aside from these two points,
$\langle N_{\rm lost} \rangle$ and $\langle E_{\rm heat} \rangle$
are smooth functions of $a$.
The number of avalanches required to get smooth results 
is particularly large in the region 
near the interval $\frac32 E_{\rm trap} < E_d < 6 E_{\rm trap}$
in which the recombination dimer is trapped 
but the recombination atom is not. 
More than 100,000 avalanches
are sometimes required to get smooth results in this region.

\subsection{Previous models}

Zaccanti et al.\ developed a simple probabilistic model 
for the avalanche process that we will refer to 
as the {\it Zaccanti model} \cite{Zaccanti:0904}.
In the Zaccanti model, the avalanche is reduced to a
discrete sequence of dimer scattering events.
A variable number of elastic collisions is followed either by the escape 
of the dimer from the trap or by a final inelastic collision.
There is one lost atom for each elastic collision up to a maximum 
number that is determined by the trap depth $E_{\rm trap}$.
The relative probability for each sequence of scattering events 
is determined by the mean column density
$\langle \int \!n\, \hbox{d}\ell \rangle$ of the trapped atoms
and by the atom-dimer cross sections $\sigma_{AD}^{\rm (el)}$ 
and $\sigma_{AD}^{\rm (in)}$.
The Zaccanti model is greatly simplified in several ways 
compared to our Monte Carlo model:
\begin{itemize}
\item
The energy dependence of 
$\sigma_{AD}^{\rm (el)}$ and $k_{\rm cm} \sigma_{AD}^{(\rm in)}$ 
is not taken into account.
These cross sections were approximated by their low-energy limits
given in Eqs.~(\ref{sigmaADel0}) and (\ref{sigmaADin0}), 
which correspond with the sharply-peaked cross sections 
in Fig.~\ref{fig:sigma}.
\item
The spacial structure of the avalanche is ignored.
All 3-body recombination events occur at the center of the cloud.
The scattering probabilities are all determined by the mean column density 
$\langle \int \!n\, \hbox{d}\ell \rangle$ averaged over directions 
from the center of the trap.
\item
The elastic scattering of the atoms is not considered.
The atom from the recombination event and the scattered atoms
from elastic atom-dimer collisions 
cannot become trapped by losing energy and they also
cannot initiate avalanches of additional lost atoms.
\item
The random variations associated with S-wave scattering
are not taken into account.
Each elastic collision decreases the kinetic energy 
of the dimer by the same multiplicative factor 5/9. 
\end{itemize}
Zaccanti et al.\ used their model to calculate the average number 
$\langle N_{\rm lost} \rangle$ of lost atoms for their experiment 
with $^{39}$K atoms \cite{Zaccanti:0904}.
It predicts that $\langle N_{\rm lost} \rangle$
increases from its background value of 3 to about 13 
near the atom-dimer resonance.  
The resulting prediction for the atom loss rate agreed 
qualitatively with the loss feature they observed near $30.4~a_0$.
The agreement could be made quantitative by decreasing 
$\sigma_{AD}^{\rm (el)}$ by a factor of 30.
Such a decrease was motivated by the 
energy dependence of the elastic atom-dimer cross section.

Machtey et al.\ developed an alternative probabilistic model 
for the avalanche process in Ref.~\cite{Khaykovich:1111}.
They made the same simplifications that were itemized above
for the Zaccanti model. 
They reduced the avalanche to discrete sequences of 
dimer scattering events whose probabilities are determined 
by an effective column density
and the atom-dimer cross sections,
but their probabilities for the sequences of scattering events
were different from the Zaccanti model. 
Another difference was that Machtey et al.\ 
never introduced the trap depth $E_{\rm trap}$ into their model.  
As a consequence, they could 
not calculate $\langle N_{\rm lost} \rangle$.
Instead they used their model to calculate 
the average number $\bar N$ of dimer collisions, 
which is not observable.
Machtey et al.\  suggested that the maximum of $\bar N$ 
as a function of $a$
might coincide with a local maximum of the atom loss rate.

\subsection{Improvements in the Model}

Our Monte Carlo model for the avalanche mechanism
has several significant improvements over the probabilistic models
proposed by Zaccanti et al.~\cite{Zaccanti:0904}
and by Machtey et al.~\cite{Khaykovich:1111}.
There are a number of further improvements that could be made. 
The approximations made in our Monte Carlo model 
are itemized at the end of Section~\ref{sec:approx}.
Many of them involve neglecting the energies of the 
low-energy atoms in the atom cloud.
One of these approximations is that the energies of the atoms 
that undergo 3-body recombination are much less than  $E_d$.
This allowed us
to use the universal rate constants 
at threshold $\alpha_{\rm shallow}$ and $\alpha_{\rm deep}$ in 
Eqs.~(\ref{alpha-shallowdeep}).
This approximation could be removed for a thermal gas 
by using the universal results for the 3-body recombination rates
in Ref.~\cite{Braaten:2008kx}, which were calculated 
up to temperatures about 100 times larger than $E_d$.
The other low-energy approximations 
allowed the trajectories of particles between collisions to be
approximated by straight lines and the changes in their 
kinetic energies between collisions to be ignored.
If the potential energies of the atoms and the dimer
(and, in the case of a BEC, their mean-field energies) 
are taken into account,
their trajectories become curved and 
their kinetic energies change between collisions
in accord with conservation of energy.
These improvements would be straightforward to implement
in our Monte Carlo model.

Another approximation is that we used a simple model 
for the trap depth that can be expressed as a change in the 
trapping potential with a single parameter $E_{\rm trap}$.  
The physics represented by that trap depth 
is actually much more complicated. 
Given our simple model for the trap depth,
it is not clear that the improvement in accuracy
from eliminating the low-energy approximations
would be worth the effort.

Finally, our rate equations for $N$ and $Q$ in Eqs.~(\ref{dNdt})
and (\ref{dQdt}) are based on the local rate equation 
for the number density in Eq.~(\ref{dndt}).
At each collision point in the avalanche, a low-energy atom is 
replaced by a high-energy atom that can then propagate through
the atom cloud.  The local approximation requires 
that most of the lost atoms come from near the recombination point
and also that most of the heat from the scattered atoms 
that are not lost is deposited near the recombination point.
Removing this local approximation would be an enormous complication.

\section{Results}
\label{sec:Results}

In this section, we apply our Monte Carlo model for the avalanche 
mechanism to the experiments with $^{39}$K and $^7$Li atoms
in which narrow loss peaks near an atom-dimer resonance 
have been observed.  We also apply it to an experiment with 
$^{133}$Cs atoms in which such a loss feature has not been observed.

\subsection{$^{39}$K atoms}

In 2009, the Florence group observed peaks in $L_3/a^4$
near $30.4~a_0$ in a BEC of $^{39}$K atoms 
and near $930~a_0$ in a thermal gas of $^{39}$K atoms \cite{Zaccanti:0904}.
Both loss features were near the predicted position of an atom-dimer resonance.
Since the van der Waals length for $^{39}$K atoms is $139~a_0$,
the loss feature near $30.4~a_0$ is in a nonuniversal region
of small scattering length.  We therefore focus on larger
scattering lengths that are safely in the universal region.
Different experimental variables were used in different regions 
of the scattering length.  We consider the experimental variables 
used in the two regions listed in Table~\ref{tab:Expvariables}.
In one region, the atom cloud was a BEC 
and in the other region, it was a thermal gas.
We choose the Efimov parameters that were determined 
from the local minimum of $L_3/a^4$ near $224~a_0$:
$a_{*} = 1140~a_0$ and $\eta_{*} = 0.043$.
For the thermal gas experiment, $E/N = 3 k_B T$ in 
Table~\ref{tab:Escales} is actually smaller than the value of
$2.7~k_B T_c$ calculated from $N$.
This suggests that the system is very close to the critical temperature,
so it might be appropriate to use the number density in 
Eq.~(\ref{ncritical}).  We nevertheless use the Boltzmann approximation 
in Eq.~(\ref{nBoltzmann}) for simplicity.

\begin{figure}[t]
\vspace*{-0.0cm}
\includegraphics*[width=0.45\linewidth,angle=0,clip=true]{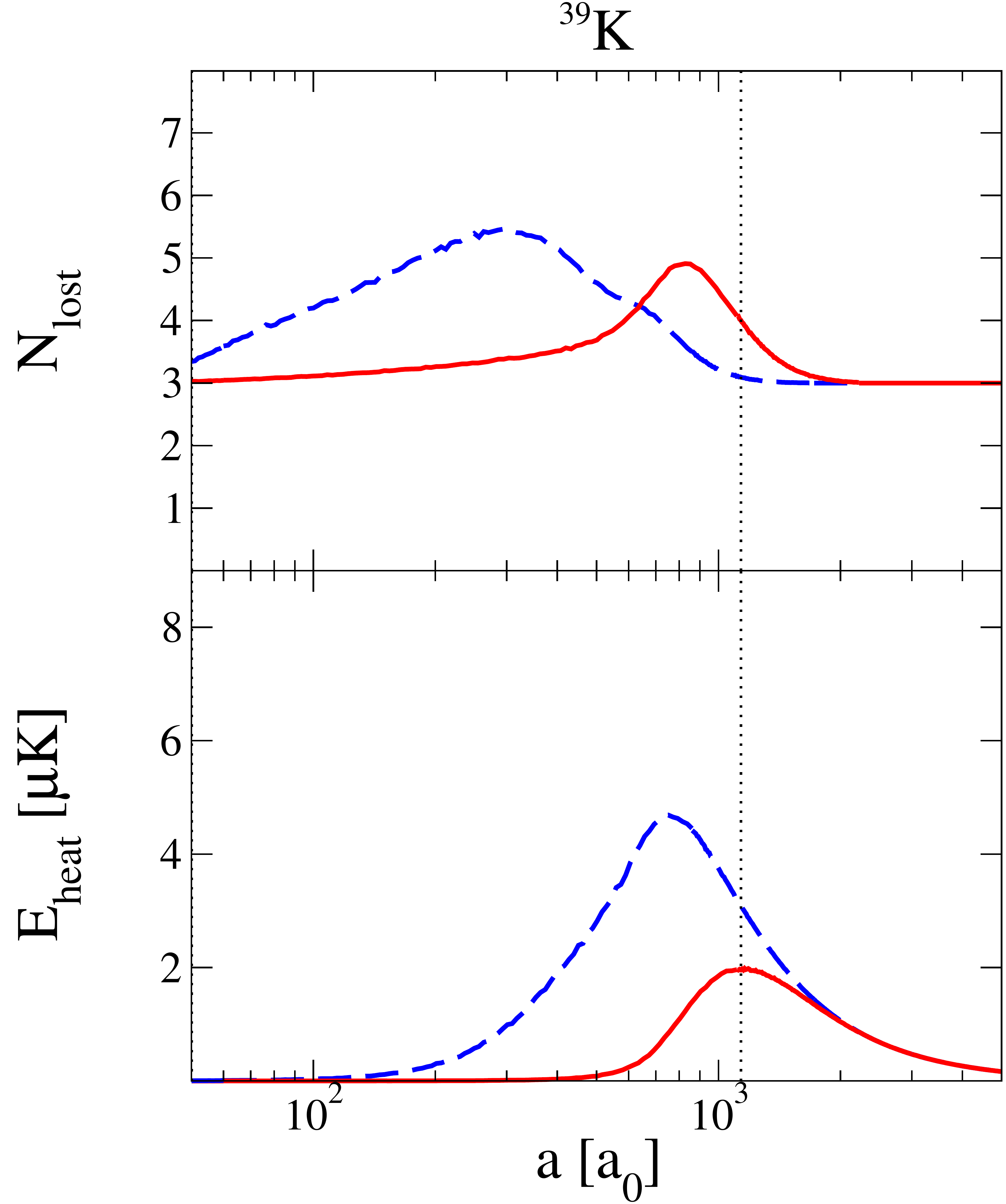}
\hspace{1cm}
\includegraphics*[width=0.45\linewidth,angle=0,clip=true]{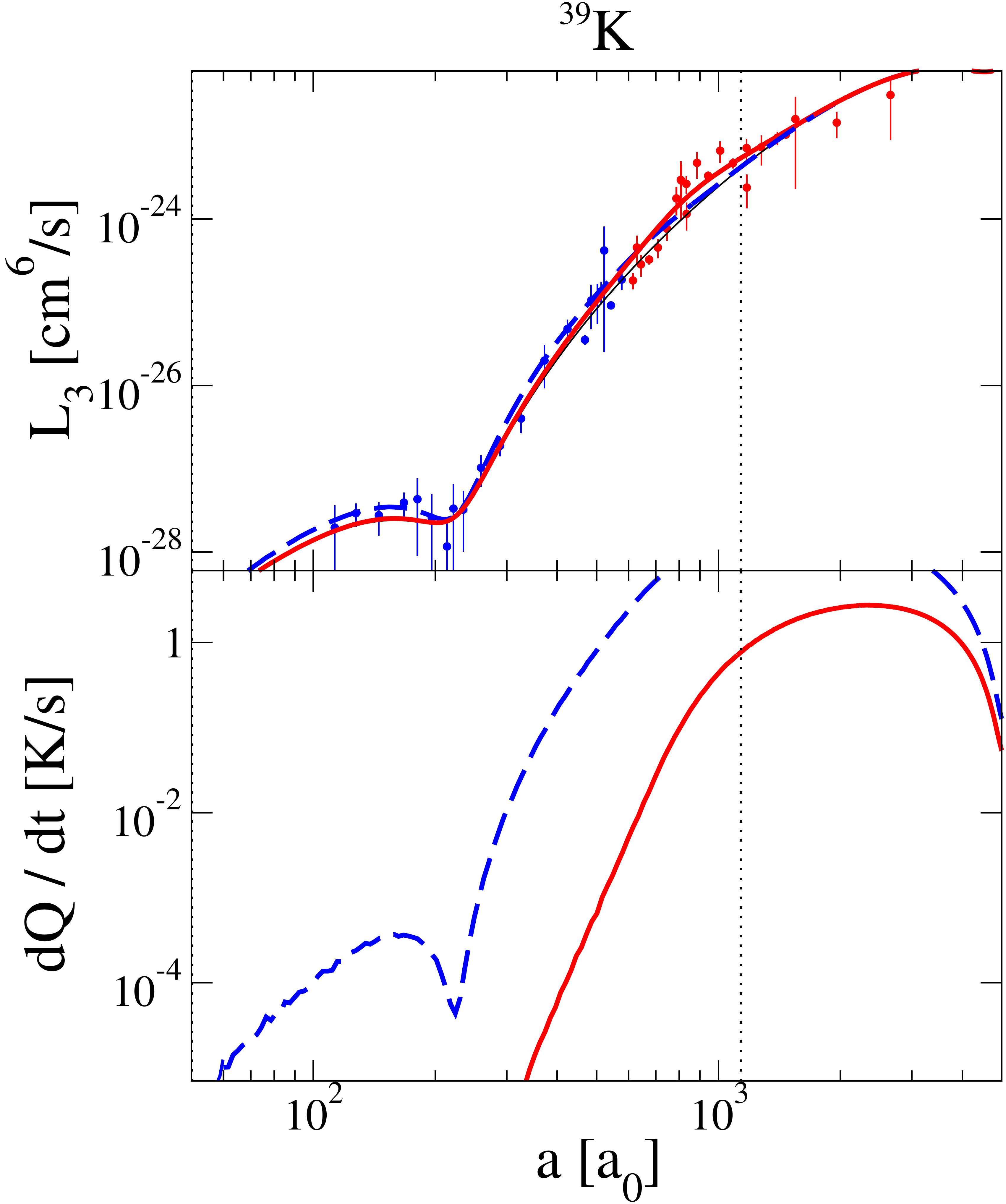}
\vspace*{-0.0cm}
\caption{(Color online) 
The average number $N_{\rm lost}$ of atoms lost in an avalanche (upper left panel),
the average heat $E_{\rm heat}$ generated by an avalanche (lower left panel),
the rate constant $L_3$ (upper right panel), and 
the heating rate $dQ/dt$ (lower right panel) as functions of $a$.
The system consists of $^{39}$K atoms
with $a_* = 1140~a_0$ and $\eta_* = 0.043$.  
The vertical dotted line marks the position of $a_*$.
The universal prediction for $L_3$ without the avalanche mechanism
is shown as a thin (black) line that provides a lower bound 
on the other curves.  Over most of the range of $a$, 
it is covered up by one of the other curves.
The dashed (blue) curves and the solid (red) curves are 
for the BEC and the thermal gas in Ref.~\cite{Zaccanti:0904}, respectively.
The data for $L_3$ are from the Florence group \cite{Zaccanti:0904}.
}
\label{fig:K39}
\end{figure}

In the left panels of Fig.~\ref{fig:K39},
the average number $N_{\rm lost}$ of atoms lost 
and the average heat $E_{\rm heat}$ from the avalanche
are shown as functions of $a$ for the two sets of experimental variables
for $^{39}$K atoms listed in Table~\ref{tab:Expvariables}.
(In this section, we omit the angular brackets that denote the avalanche
averages of $N_{\rm lost}$ and $E_{\rm heat}$.)
For both the BEC and the thermal gas, 
$N_{\rm lost}$ and $E_{\rm heat}$ are shown for a 
range of scattering lengths that extend over two orders of magnitude.
The rate constant $L_3$ however was measured  
using these experimental variables only over the smaller ranges
of $a$ specified in Table~\ref{tab:Expvariables}.
For both the BEC and the thermal gas, $N_{\rm lost}$ has a broad peak 
with a maximum value near 5.  The position of the peak is at $293~a_0$ 
for the BEC and at $827~a_0$ for the thermal gas.
This position is determined by the atom-dimer cross sections
and the trap depth, among other things. 
As illustrated in Fig.~\ref{fig:sigma},
the cross section for the first elastic scattering
of the recombination dimer has a broad peak with maximum at 
$4.34 (e^{-\pi/s_0} a_*) \approx 218~a_0$.
The trap depth forces $N_{\rm lost}$ to decrease to the naive value 3 
when $E_d = \frac32 E_{\rm trap}$, which is near 
$a=1720~a_0$ for the BEC and near $a=2220~a_0$ for the thermal gas.
Thus the cutoff provided by the trap depth has a strong effect 
on the position of the peaks in $N_{\rm lost}$ and $E_{\rm heat}$.

In the right panels of Fig.~\ref{fig:K39},
the rate constant $L_3$ and the heating rate $dQ/dt$
are shown as functions of $a$.
The panel for $L_3$ in Fig.~\ref{fig:K39} shows the data 
from the Florence group \cite{Zaccanti:0904}.
The result for $L_3$ is visibly larger than the universal 
result without the avalanche mechanism in the region just below the
local minimum near $e^{-\pi/s_0}a_+ = 225~a_0$ for the BEC
and in the region just below $a_*= 1140~a_0$ for the thermal gas.
Thus the avalanche mechanism can affect the fitted values 
of the Efimov parameters $a_*$ and $\eta_*$.
Both $L_3$ and $dQ/dt$ have local minima near 
$e^{-\pi/s_0}a_+ = 225~a_0$ that arise from Efimov interference.
The heating rate $dQ/dt$ in the thermal gas is more than an order of magnitude 
smaller than in the BEC over most of the range of $a$.

\subsection{$^7$Li atoms}

\begin{figure}[t]
 \vspace*{-0.0cm}
\includegraphics*[width=0.45\linewidth,angle=0,clip=true]{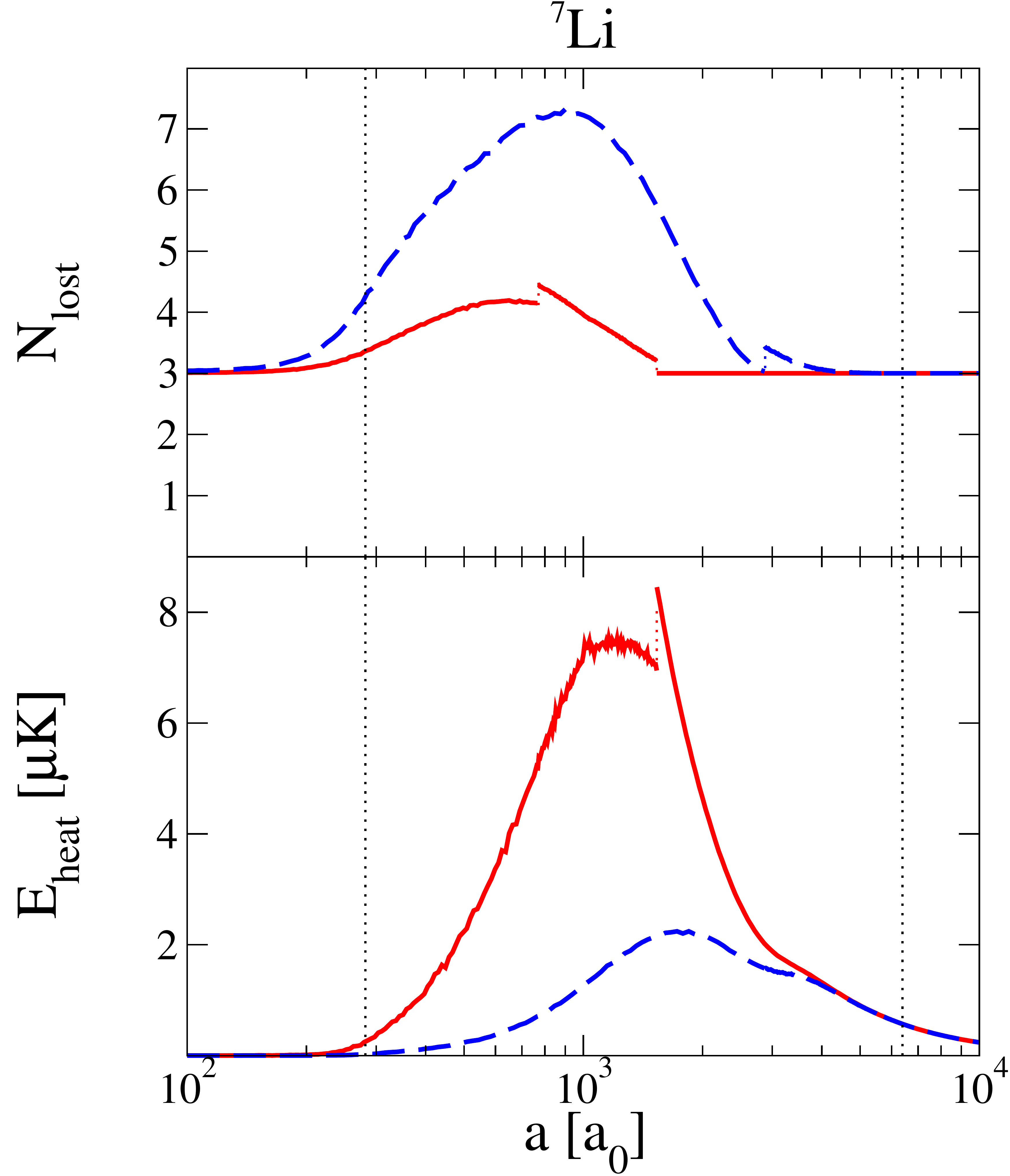}
\hspace{1cm}
\includegraphics*[width=0.45\linewidth,angle=0,clip=true]{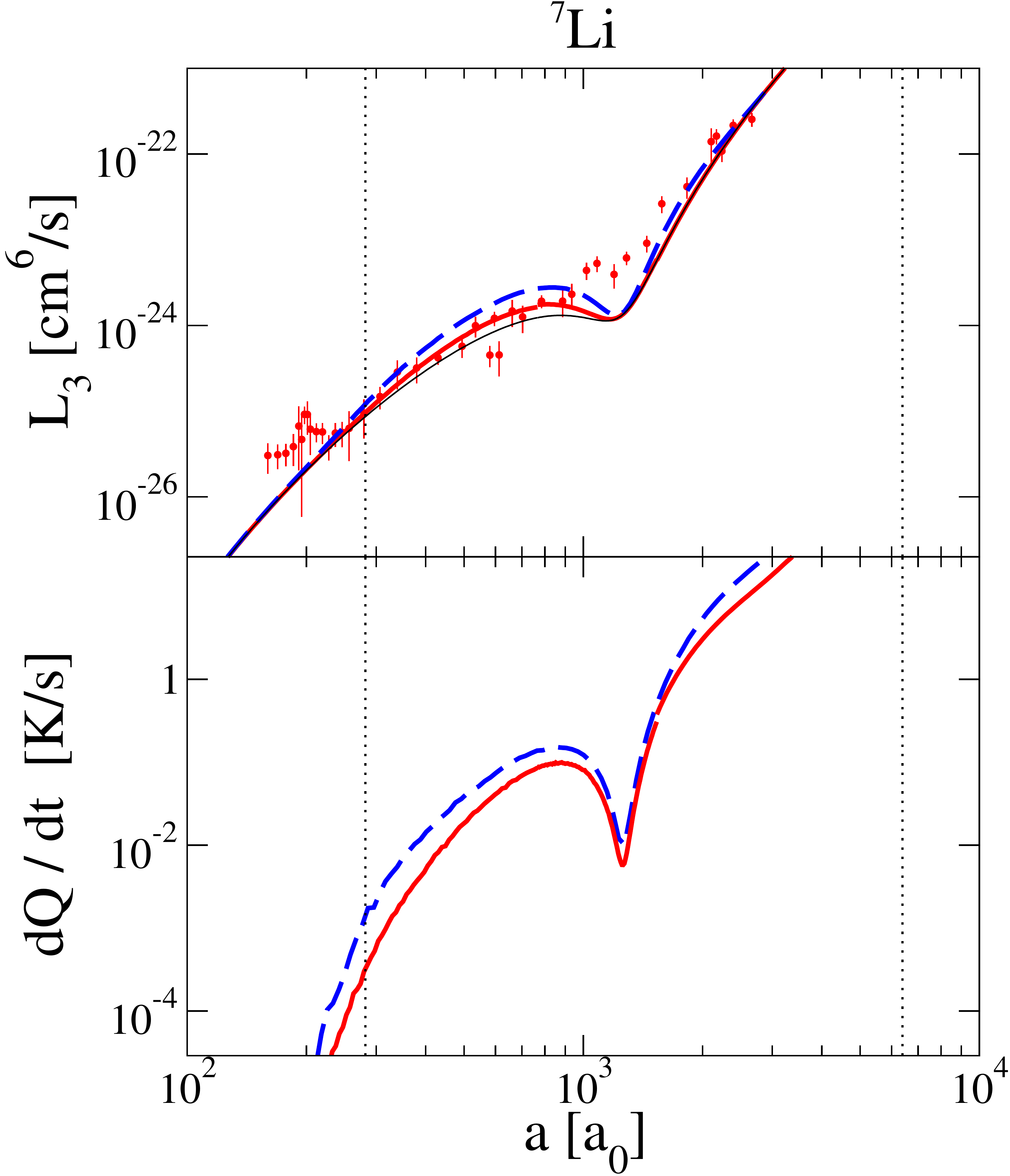}
\vspace*{-0.0cm}
\caption{(Color online)  
Same as Fig.~\ref{fig:K39},
but for $^7$Li atoms
with $a_* = 282~a_0$ and $\eta_* = 0.039$.  
The dashed (blue) curves and the solid (red) curves are 
for the BEC in Ref.~\cite{Hulet:0911} 
and the thermal gas in Ref.~\cite{Khaykovich:1201}, respectively.
The data for $L_3$ are from the Bar-Ilan group \cite{Khaykovich:1201}.
}
\label{fig:Li7}
\end{figure}

In 2009, the Rice group observed a peak in $L_3/a^4$
near $608~a_0$ in a BEC of $^{7}$Li atoms 
in the $|1,+1\rangle$ hyperfine state \cite{Hulet:0911}.
This loss feature is near the predicted position of an atom-dimer resonance.
Different experimental variables were used in different regions 
of the scattering length.  The experimental variables 
used in one of these regions are listed in Table~\ref{tab:Expvariables}.
The Efimov parameters determined 
from the narrow loss minimum near $2676~a_0$
are $a_{*} = 598~a_0$ and $\eta_{+} = 0.039$.
The Rice group has improved the accuracy of the determination of $a$ 
as a function of the magnetic field
and a reanalysis of the data from Ref.~\cite{Hulet:0911}
is underway \cite{Hulet:1205}.
Their new analysis will not have a significant effect on the value of 
$\eta_*$, but it will shift the value of $a_*$ 
closer to the value measured by the Bar-Ilan group, which is given below.

In 2010, the Bar-Ilan group observed a local minimum in $L_3/a^4$
in a thermal gas of $^{7}$Li atoms 
in the $|1,+1\rangle$ hyperfine state \cite{Khaykovich:1003}.
The Efimov parameters determined by fitting their measurements of $L_3$
are $a_{*} = 282~a_0$ and $\eta_{*} = 0.188$.
Since the van der Waals length for $^{7}$Li atoms is $65~a_0$,
the predicted atom-dimer resonance is safely in a universal region 
of large scattering length.
In 2012, they presented additional data that revealed
a narrow enhancement in $L_3$ near $200~a_0$,
which is near the predicted atom-dimer resonance \cite{Khaykovich:1201}.
The experimental variables used in the measurement of $L_3$ 
are listed in Table~\ref{tab:Expvariables}.

The Rice group and the Bar-Ilan group used the same hyperfine 
state of $^{7}$Li, so they should obtain the same Efimov parameters 
to within experimental errors.  Since $\eta_*$ is particularly sensitive 
to the width of the loss minimum at $a_+$, thermal smearing and 
limited experimental resolution 
are most likely to lead to an overestimate of $\eta_*$.
For the Efimov parameters, we will therefore use the value of $a_*$ 
obtained by the Bar-Ilan group but the smaller value of $\eta_*$ 
obtained by the Rice group:  $a_{*} = 282~a_0$ and $\eta_{*} = 0.039$.

In the left panels of Fig.~\ref{fig:Li7},
the average number $N_{\rm lost}$ of atoms lost 
and the average heat $E_{\rm heat}$ from the avalanche
are shown as functions of $a$ for the two sets of experimental variables
for $^{7}$Li atoms listed in Table~\ref{tab:Expvariables}.
There are visible discontinuities in $N_{\rm lost}$ and $E_{\rm heat}$
at the scattering lengths at which
$E_d = \frac32 E_{\rm trap}$ and $E_d = 6 E_{\rm trap}$.
These discontinuities are artifacts of our simple model for 
the trap depth.  
The number $N_{\rm lost}$ has a broad peak 
with a maximum value near 7 for the BEC and near 4 for the thermal gas.  
The position of the peak is at $905~a_0$ 
for the BEC and at $767~a_0$ for the thermal gas.
This position is determined the scattering cross sections 
and the trap depth, among other things.  
The cross section for the first elastic scattering
of the recombination dimer has a broad peak with its maximum at 
$4.34 a_* \approx 1220~a_0$.
The trap depth forces $N_{\rm lost}$ to decrease to the naive value 3 near 
$a=5740~a_0$ for the BEC and near $a=1530~a_0$ for the thermal gas.
This cutoff provided by the trap depth has a strong effect 
on the position of the peaks in $N_{\rm lost}$ and $E_{\rm heat}$.

In the right panels of Fig.~\ref{fig:Li7},
the rate constant $L_3$ and the heating rate $dQ/dt$
are shown as functions of $a$.
The panel for $L_3$ in Fig.~\ref{fig:Li7}
shows the data from the Bar-Ilan group \cite{Khaykovich:1003}.
The curves for $L_3$ have a much more pronounced local minimum at 
$a_+ \approx 1260~a_0$ than the data, because we have used the 
Efimov parameter $\eta_* = 0.039$ from the Rice experiment \cite{Hulet:0911}
instead of the value $\eta_* = 0.188$ obtained by fitting the Bar-Ilan data.
For both the BEC and the thermal gas, the result for $L_3$ 
in the region just below the local minimum near $a_+ = 1260~a_0$
is visibly larger than the universal 
result without the avalanche mechanism.
Since the Efimov parameters $a_+$ and $\eta_*$ are sensitive to the
position and width of the minimum, their fitted values 
can be strongly affected by the avalanche mechanism.
Note that our Monte Carlo model predicts no peak in $L_3$ 
near the atom-dimer resonance at $a_* = 282~a_0$., 
Both $L_3$ and $dQ/dt$ have local minima near 
$a_+ = 1260~a_0$ that arise from Efimov interference.
The heating rates $dQ/dt$ are similar in the BEC and in the thermal gas.

\subsection{$^{133}$Cs atoms}

\begin{figure}[t]
\vspace*{-0.0cm}
\includegraphics*[width=0.45\linewidth,angle=0,clip=true]{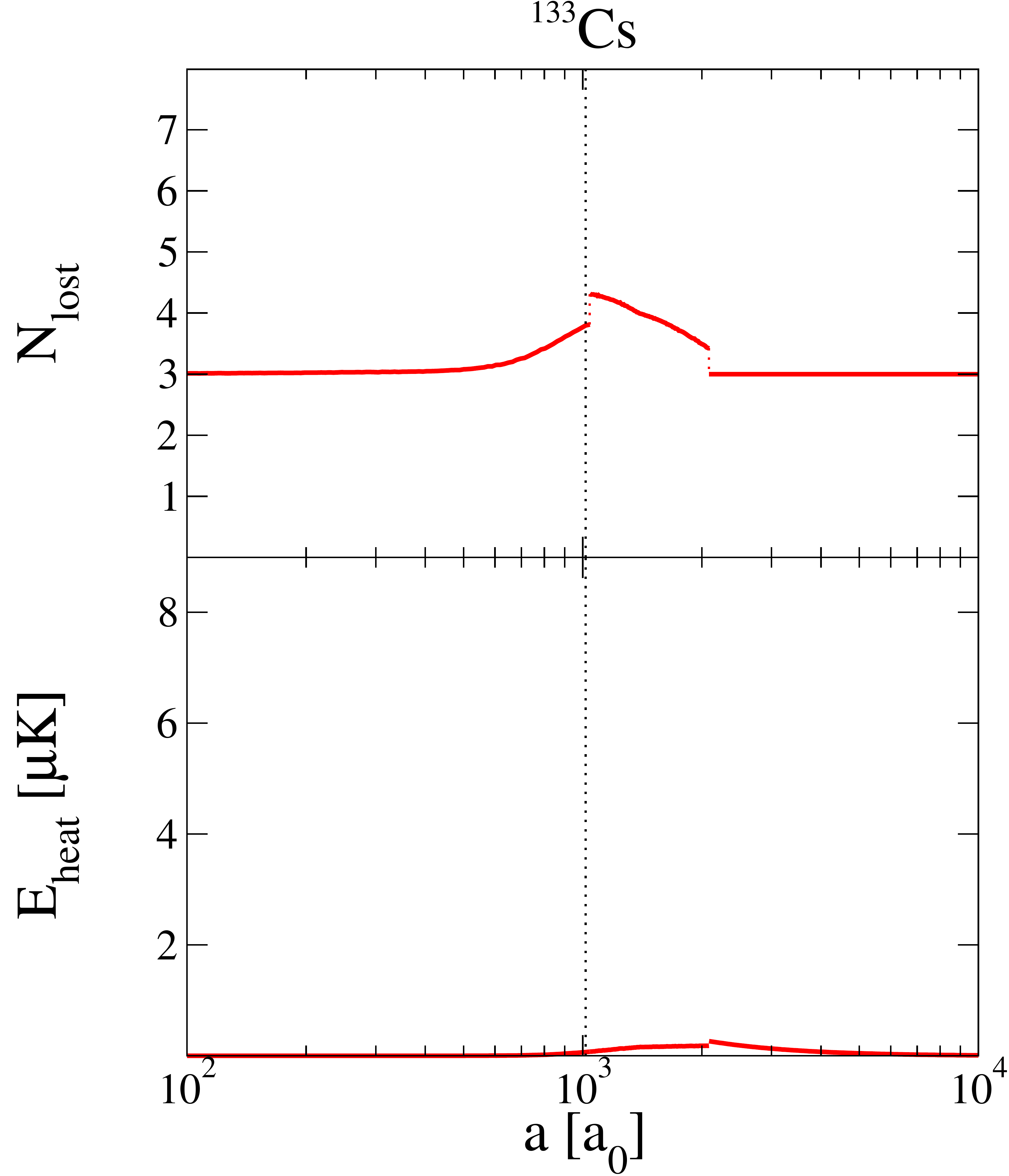}
\hspace{1cm}
\includegraphics*[width=0.45\linewidth,angle=0,clip=true]{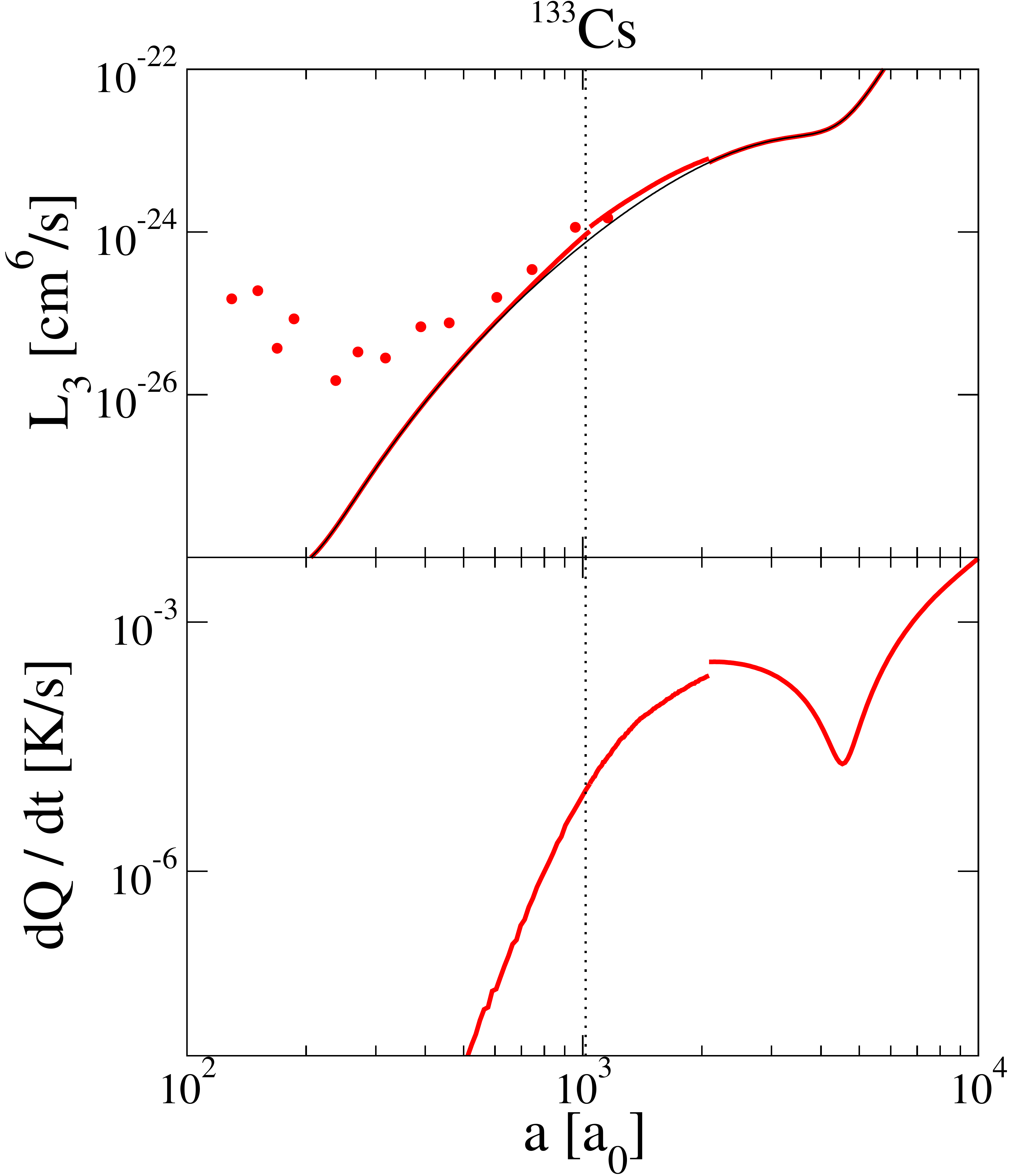}
\vspace*{-0.0cm}
\caption{(Color online) 
Same as Fig.~\ref{fig:K39},
but for $^{133}$Cs atoms
with $a_* = 1017~a_0$ and $\eta_* = 0.08$.  
The solid (red) curves are 
for the thermal gas in Ref.~\cite{Grimm:1106}.
The data for $L_3$ are from Ref.~\cite{Grimm:1106}.
}
\label{fig:Cs133}
\end{figure}

The Innsbruck group has studied loss features in thermal gases of 
$^{133}$Cs atoms in several universal regions of the magnetic field.
In 2005, they studied a region of low magnetic field
and observed a local minimum of $L_3/a^4$ that can be 
attributed to Efimov interference 
at a scattering length near $210~a_0$ \cite{Grimm:0512}.  
Setting $a_+ = 210~a_0$, atom-dimer resonances are 
predicted at $47~a_0$ and $1070~a_0$.  
Since the van der Waals length for $^{133}$Cs atoms is $200~a_0$,
only the higher atom-dimer resonance is in a universal region 
of large scattering length.  The Innsbruck group 
did not observe any loss feature near $1070~a_0$.

In 2011, the Innsbruck group observed three peaks in $L_3/a^4$
in different universal regions 
with large negative $a$ at higher magnetic field \cite{Grimm:1106}.
They can all be attributed to resonant enhancement from
an Efimov trimer near the 3-atom threshold.
Two of these loss features are complicated by the presence of a 
G-wave Feshbach resonance.
The Efimov parameters associated with the third loss feature are
$a_- = -955~a_0$ and $\eta_* = 0.08$.
The Innsbruck group found that the Efimov parameter $a_-$ 
has almost the same value in all the universal regions, which suggests 
that it is determined by the van der Waals length \cite{Grimm:1106}.
The universal ratios in Eqs.~(\ref{eq:a*-a*}) and (\ref{eq:a*0a*})
can be used to predict an atom-dimer resonance 
at $a_* = 1017~a_0$ and an Efimov interference minimum 
in $L_3/a^4$ near $a_+ = 200~a_0$.
The Innsbruck group measured $L_3$ in a universal region 
of large positive $a$.  They
observed a local minimum near $270~a_0$, 
which is near the predicted position of $a_+$,
but they did not see any loss feature near $a_*$.
The experimental variables used in this region of 
positive scattering length are listed in Table~\ref{tab:Expvariables}.

In the left panels of Fig.~\ref{fig:Cs133},
the average number $N_{\rm lost}$ of atoms lost 
and the average heat $E_{\rm heat}$ from the avalanche
are shown as functions of $a$ for this set of experimental variables.
The number of lost atoms coincidentally has a peak 
very close to the atom-dimer resonance $a_*$
but the peak in $E_{\rm heat}$ is at a higher value of $a$.
The average heat $E_{\rm heat}$ is more than an order of magnitude smaller 
than in the $^7$Li and $^{39}$K thermal gas experiments described above.

In the right panels of Fig.~\ref{fig:Cs133},
the rate constant $L_3$ and the heating rate $dQ/dt$
are shown as functions of $a$.
The increase in $L_3$ from the avalanche mechanism is visible 
only in the region just above $a_* = 1017~a_0$.
Both $L_3$ and $dQ/dt$ have local minima
from Efimov interference near $a_+=4550~a_0$.
The panel for $L_3$ in Fig.~\ref{fig:Cs133}
also shows the data from the Innsbruck group \cite{Grimm:1106}.
There is a local minimum near $200~a_0$, which is near the predicted 
value of $a_+$.  
The deviations between the measurements and the universal predictions 
at smaller $a$ is not unexpected,
because this is a nonuniversal region.

\section{Dimer-dimer resonances}
\label{sec:Dimer-dimer}

We have used our Monte Carlo model for the avalanche mechanism 
to show that it cannot produce a narrow loss feature 
near an atom-dimer resonance.
The essential reason is explained by the energy dependence 
of the elastic atom-dimer cross section, which is illustrated in
Fig.~\ref{fig:sigma}.  Since the shallow dimer from 3-body recombination 
loses energy with each elastic collision, the first few collisions 
of the dimer are those that are the most likely to knock an extra atom 
out of the trap.
In the first few collisions, the dimer's kinetic energy is comparable to 
its binding energy $E_d$, and there is no dramatic enhancement 
of the atom-dimer cross section.
Rather, the atom-dimer cross section is dramaticly enhanced only 
after many elastic collisions have reduced the dimer's kinetic energy 
to much smaller than $E_d$. However, with its kinetic energy degraded,
the dimer is much less likely to knock an atom out of the trap.

The universality of atoms with large scattering length
is not limited to the 2-atom and 3-atom sectors.
In 2004, Hammer, Meissner, and Platter made the first suggestion
that universality should extend to the 4-atom sector \cite{Platter:2004qn}.
They presented numerical evidence that there are two universal 
tetramers associated with each Efimov trimer, and they made the 
first calculations of the tetramer binding energies 
for some regions of $1/a$ \cite{Platter:2004qn,Hammer:2006ct}.
In 2008, von Stecher, D'Incao, and Greene calculated the
tetramer binding energies 
more accurately and over the entire range of $1/a$ \cite{vSIG:0810}.
They pointed out that the most dramatic signature of a universal tetramer
is the resonant enhancement of the 4-body recombination rate 
at a negative value of $a$ where the tetramer is at the 4-atom threshold.
The loss features from a pair of universal tetramers were
first observed by the Innsbruck group using a thermal gas of 
$^{133}$Cs atoms \cite{Grimm:0903}.
They measured the 4-body loss rate contant $L_4$ 
for one of the tetramers.
Universal tetramers have also been observed in a thermal gas
experiment with $^7$Li atoms by the Rice group \cite{Hulet:0911}.
They measured $L_4$ for both members of one pair of tetramers
and for one member of another pair.

A universal tetramer could also produce loss features
at positive values of $a$.
One possibility is a loss feature at a scattering length
at which a tetramer crosses the 2-dimer threshold,
which we will refer to as a {\it dimer-dimer resonance}.
The dimer-dimer elastic and inelastic cross sections 
are resonantly enhanced near threshold at a dimer-dimer resonance.
Each Efimov trimer, with atom-dimer resonance at $a_{*}$, 
has associated with it two universal tetramers, with dimer-dimer 
resonances at larger scattering lengths $a_{*1}$ and $a_{*2}$.
The universal predictions for the positions 
of these dimer-dimer resonances were first calculated 
by von Stecher, D'Incao, and Greene \cite{vSIG:0810}.
They were recently calculated by Deltuva 
with 4 digits of precision \cite{Deltuva:2011ur}:
\begin{subequations}
\begin{eqnarray}
a_{*1} &=& 2.196~a_*.
\label{eq:a*1a*}
\\
a_{*2} &=& 6.785~a_*.
\label{eq:a*2a*}
\end{eqnarray}
\end{subequations}

In their experiment with a BEC of $^7$Li atoms, the Rice group
observed narrow enhancements in the measured 
3-body loss rate constant $L_3$
near $+1470~a_0$ and near $+3910~a_0$ \cite{Hulet:0911}.
These scattering lengths are near the predicted positions 
of the two dimer-dimer resonances for a pair of universal tetramers.
These loss features are even more mysterious than those
near atom-dimer resonances.
Three-body recombination can create a shallow dimer with kinetic energy
comparable to $E_d$.
Four-body recombination can create one or two shallow dimers
with kinetic energy comparable to $E_d$.
If the equilibrium population of shallow dimers 
in the atom cloud is negligible,
the recombination dimers can only undergo atom-dimer collisions. 
Thus the resonant enhancement of dimer-dimer cross sections 
near a dimer-dimer resonance 
should have no effect on the atom loss rate.
Therefore, there is no analog of the avalanche mechanism 
near a dimer-dimer resonance.  

One possible explanation for the narrow loss features near the 
atom-dimer and dimer-dimer resonances is that the equilibrium population 
of shallow dimers in the atom cloud is not negligible.
The resonant enhancement of the inelastic atom-dimer 
cross section could then produce an enhanced loss rate 
near $a_*$.  Similarly,
the resonant enhancement of the inelastic dimer-dimer cross section 
could produce enhanced loss rates near $a_{*1}$ and near $a_{*2}$.
The number density $n_d$ of the dimers must be much 
smaller than the number density $n$ of the atoms.  
In the absence of atom loss processes,
the rates of atom-dimer and dimer-dimer collisions 
would be proportional to $n n_d$ and $n_d^2$, respectively.
If $n_d$ is proportional to $n^2$,
the atom-dimer and dimer-dimer collision rates have the same 
dependence on $n$ as the 3-body and 4-body recombination rates,
respectively. 
Thus the enhanced loss rate near $a_*$ from low-energy inelastic
atom-dimer collisions would manifest itself 
as an apparent enhancement of the 3-body recombination rate.
Similarly, enhanced loss rates near $a_{*1}$ and near $a_{*2}$ 
from low-energy inelastic atom-dimer collisions would manifest themselves 
as apparent enhancements in the 4-body recombination rate.
While an equilibrium population of dimers
could explain the existence of  loss features at the 
atom-dimer and dimer-dimer resonances,
it can not easily account for them quantitatively.
It seems likely that a population of dimers 
large enough to account for the observed loss features
should also have been observed more directly.

Narrow loss features have been observed near atom-dimer resonances 
in several experiments and near dimer-dimer resonances 
in the $^7$Li BEC experiment.
We have shown that the avalanche mechanism cannot produce 
a narrow loss feature near an atom-dimer resonance.
It also cannot produce any loss of atoms near a dimer-dimer resonance.
An equilibrium population of dimers
could produce loss features near atom-dimer and dimer-dimer resonances,
but a population of dimers large enough to account for the observed loss
features should probabily have been observed more directly.
We suggest that another mechanism that has not yet been identified
must be responsible for the loss features that have been observed
near atom-dimer and dimer-dimer resonances.

\begin{acknowledgments}
We thank R.~Grimm, R.~Hulet, L.~Khaykovich, H.-C.~N\"agerl, and M.~Zaccanti 
for useful communications.
We also thank M.~Zaccanti, L.~Khaykovich, and M.~Berninger and A.~Zenesini 
for providing their data.
This research was supported in part by a joint grant from 
the Army Research Office 
and the Air Force Office of Scientific Research.
\end{acknowledgments}

\end{document}